\documentclass[a4paper,11pt]{article}
\usepackage{jinstpub} 
\usepackage{subcaption}
\usepackage{graphicx}
\usepackage{placeins}
\usepackage{lineno}
\usepackage{natbib}

\title{\boldmath The CONDOR Observatory: A Gamma-Ray Observatory with a 100 GeV Threshold at 5300 Meters Above Sea Level}

\author[c]{Miguel Arratia}
\author[a,d]{,W.K. Brooks}
\author[e]{Jiajun Huang}
\author[a]{,Gonzalo Muñoz J.}
\author[d]{,Luis Navarro F.}
\author[c]{,Sebouh J. Paul}
\author[b,d]{,Raquel Pezoa R.}
\author[a,d]{,Sebastian Tapia}
\author[a]{,Daniel Torres A. }
\author[a,e,f]{,Constanza Valdivieso C.}
\author[a,e]{,Nicolas Viaux M.}

 \affiliation[a]{Departamento de Física, Universidad Técnica Federico Santa María, Valparaíso, Chile}
 \affiliation[b]{Departamento de Informática, Universidad Técnica Federico Santa María, Valparaíso; Chile}
 \affiliation[c]{Department of Physics and Astronomy, University of California, Riverside, USA}
 \affiliation[d]{Centro Cientifico Tecnólogico de Valparaíso (CCTVal), Universidad Técnica Federico Santa María, Valparaíso, Chile }
\affiliation[e]{Millennium Institute for Subatomic physics at high energy frontier (SAPHIR), Santiago, Chile}
\affiliation[f]{Instituto de Física, Pontificia Universidad Católica de Valparaíso, Avenida Universidad 331, Curauma, Valparaíso, Chile}

\emailAdd{nicolas.viaux@usm.cl}

\abstract{We present the design of the COmpact Network of Detectors with Orbital Range (CONDOR), a proposed high-altitude gamma-ray and cosmic-ray (CR) observatory set to become the highest of its kind. Planned for installation at Cerro Toco in the Atacama Desert, Chile, at 5300 meters above sea level (m.a.s.l.), CONDOR is optimized to operate in the 100 GeV to 1 TeV range using the extensive air-shower technique. The design prioritizes simplicity, modularity, and robustness to ensure reliable performance in a harsh environment. The CONDOR array has a fill factor of 90\% and consists of 6000 plastic scintillator panels, each approximately 1 m$^{2}$, read out by wavelength-shifting fibers and SiPMs. The readout electronics are based on flash ADCs, with White Rabbit technology ensuring time synchronization. We present an analysis of angular reconstruction and particle discrimination using CORSIKA-simulated CR showers, developing methods to reconstruct incident angles and distinguish between gamma-ray and proton CR 
events. CONDOR’s exceptional altitude and compact design enable it to meet the target 100 GeV threshold, complementing other ground-based observatories and overlapping with satellite detection ranges. CONDOR has the potential to support an extensive research program in astroparticle physics and multimessenger astronomy from the Southern Hemisphere, operating in all‐sky mode 24/7.}


\arxivnumber{} 

\begin{document}
\maketitle
\flushbottom

\section{Introduction}
\label{sec:intro}
Cosmic rays (CR) provide a unique window into the most energetic processes of the Universe. Upon entering Earth's atmosphere, CR whether charged particles or gamma rays—interact with air molecules, generating extensive air showers (EAS) of secondary particles such as pions and muons that enable their indirect detection. These studies are essential for addressing fundamental questions about the origins, acceleration mechanisms, and propagation of CR through the interstellar medium.

The detection techniques for cosmic vary across their energy spectrum. For low energies ($10^9 - 10^{14} , \text{eV}$), direct measurements of charged CR using 
satellites and high-altitude balloon experiments, like PAMELA, AMS-02, and CREAM~\cite{picozza2007pamela, kirn2013ams, ahn2008measurements}, provide 
detailed insights into fluxes and composition. Similarly, satellite experiments such as Fermi-LAT~\cite{Atwood_2009} were designed to measure gamma rays up to a few hundred GeV. At higher energies, where the flux decreases significantly, ground-based detectors become indispensable. These instruments measure secondary particles produced in EAS through technologies such as scintillator arrays, water Cherenkov tanks, Cherenkov cameras, radio antennas, etc.

Several observatories have been designed to measure CR at high altitudes using the EAS technique~\cite{DiSciascio:2019lse}. For example, the ARGO-YBJ observatory~\cite{Argo-YBJ} operated until 2013, was located at an altitude of 4,300 m.a.s.l. in Tibet, China, with an energy range of a few hundred GeV to a few TeV and a physical scale of 11,000 m$^{2}$. The HAWC observatory~\cite{HAWK}, located on the Sierra Negra volcano in Mexico at 4,100 m.a.s.l., is currently operational with an energy range from several hundred GeV to several hundred TeV, covering an area of 22,000 m$^{2}$. 

The ALPACA experiment~\cite{ALPACA}, under construction near Chacaltaya, Bolivia, at approximately 4,740 m.a.s.l., is designed for an energy range from a few TeV to PeV and will cover an area of 83,000 m$^{2}$. The LHAASO observatory~\cite{LHAASO}, operating in Daocheng, China, at 4,410 m.a.s.l., has an energy range from a few hundred GeV to PeV and extends over 1 km$^{2}$. A future addition to this class of observatories is the SWGO project~\cite{SWGO}, planned for Pampa La Bola in Atacama, Chile, at an altitude of 4,770 m.a.s.l., with an energy range of a few hundred GeV to a few PeV and a proposed coverage area of 200,000 m$^{2}$.

The Southern Hemisphere offers a unique observational advantage for CR studies, particularly in regions such as the Galactic Center and the Magellanic Clouds. These conditions allow for an in-depth investigation of the interplay between galactic magnetic fields and CR propagation, especially in the hundreds-of-GeV energy range.

The mean number of electrons at ground level strongly depends on atmospheric depth and primary gamma-ray energy. Placing a ground-based gamma-ray observatory at high altitude enhances detection efficiency, especially at lower energies, improving its ability to differentiate between primary CR components~\cite{DiSciascio:2019lse,DiSciascio:2022hyc}.

\begin{figure}[ht!]
\centering
\includegraphics[width=0.5\linewidth]{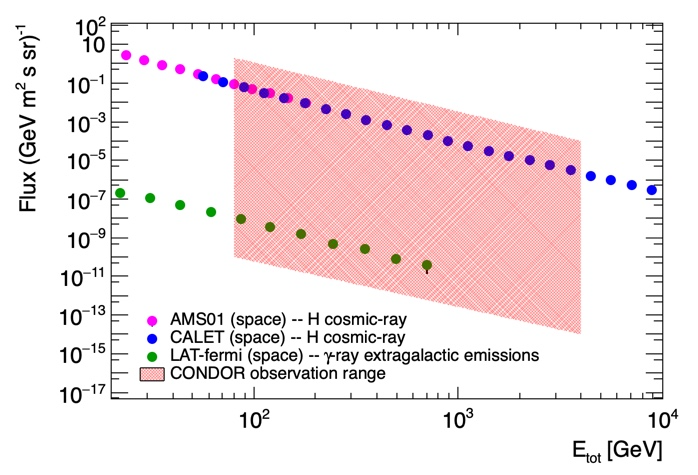}
\caption{Measurements of CR flux from LAT-fermi, CALET and AMS. In red the region of interest for the CONDOR observatory.}
\label{fig:Flux}
\end{figure}

Figure~\ref{fig:Flux} presents measurements of CR flux from LAT-fermi~\cite{LAT-fermi}, CALET~\cite{CALET2022} and AMS~\cite{AMS01} together with the expected range of observation of CONDOR. This establishes the requirements for background rejection in gamma-ray measurements in this range.  

This study represents the first phase in the development of the CONDOR Observatory. We present a conceptual design, the results of an in-situ test of key components, demonstrate the feasibility of angular reconstruction techniques and gamma-proton identification.

\section{The CONDOR Observatory}
\label{sec2}
The CONDOR Observatory is uniquely positioned to observe extensive air showers at lower atmospheric depths, where the number of secondary electrons remains high. It consists of 5,300 plastic scintillator detectors arranged in a central array, with an additional 1,040 detectors forming a peripheral array. This configuration enables precise angular reconstruction and differentiation of primary particles such as protons and gamma rays in the 100 GeV--1 TeV range.

\begin{figure}[ht!]
\centering
\includegraphics[width=\linewidth]{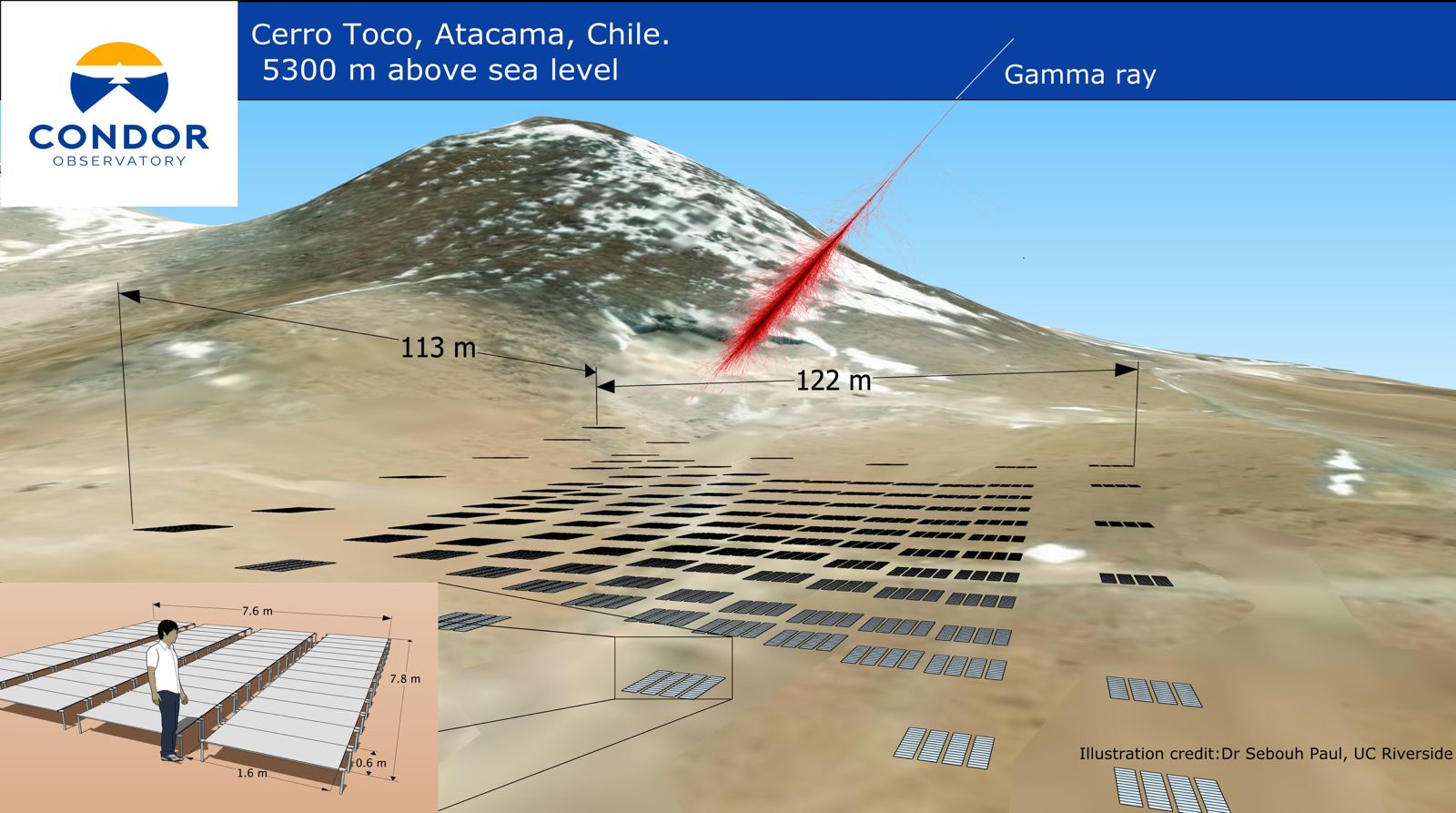}
\caption{Schematic representation of the CONDOR Observatory.}
\label{fig:schematic_represantation}
\end{figure}

The CONDOR array will be located at the Atacama Astronomical Park~\cite{PAA}, situated on the Chajnantor Plateau in Atacama, Chile. The array is designed to be compact, achieving a fill factor exceeding 90\% in its central detector—which covers an area of approximately 6,000 $\text{m}^2$—and an outer veto region, all enclosed within a total area of 14,000 $\text{m}^2$. A schematic representation of the CONDOR array, illustrating its compact layout and the implementation of the veto region, is shown in Figure~\ref{fig:schematic_represantation}. 

\begin{figure}[ht!]
    \centering
    \begin{subfigure}{0.495\linewidth}
        \centering
        \includegraphics[width=0.96\linewidth]{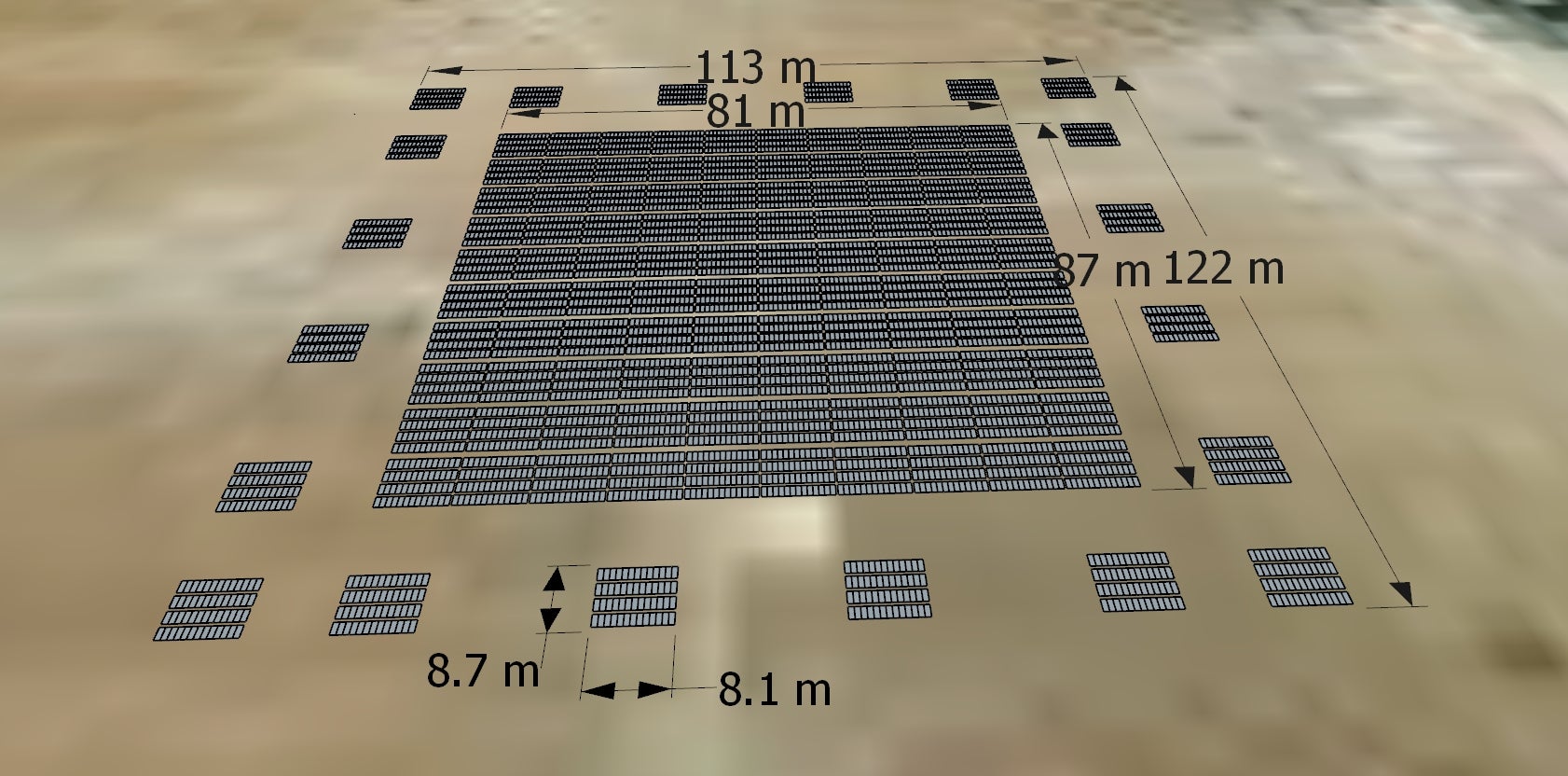}
        \caption{}
        \label{fig:subfig1}
    \end{subfigure}
    \hfill
    \begin{subfigure}{0.495\linewidth}
        \centering
        \includegraphics[width=0.855\linewidth]{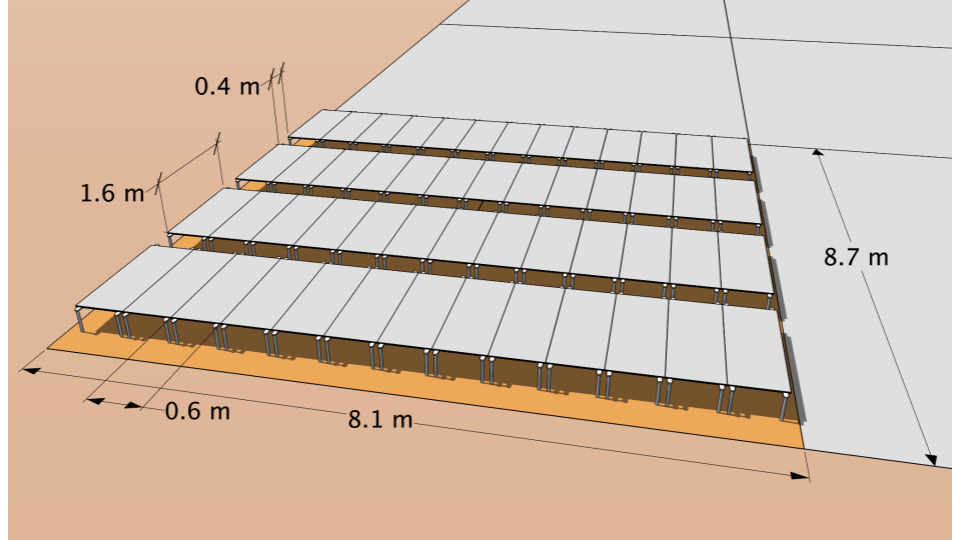}
        \caption{}
        \label{fig:subfig2}
    \end{subfigure}
\caption{(a) CONDOR array layout, (b) zoom of a cell unit}
\label{fig:layout}
\end{figure}

The central array primarily captures the core of the EAS, while the peripheral detectors are part of its veto system to effectively exclude background events, thereby enhancing the accuracy of gamma-ray observations. This configuration minimizes gaps between the panels, allowing just enough space for servicing and maintenance. The outer veto region follows a similar implementation. This layout, inspired by the ARGO-YBJ experiment~\cite{Argo-YBJ} and presented in Figure~\ref{fig:layout}, is optimized for low-energy gamma-ray detection, targeting energies as low as 100 GeV. Note that ARGO-YBJ had a threshold of about 300 GeV, employing a similar layout and Resistive Plate Chambers technology, with sensitivity comparable to that of plastic scintillators for primarily charged particles. Thanks to its significantly higher altitude, CONDOR will lower this threshold. Thus, CONDOR adopts an approach similar to that advocated by DiSciascio~\cite{DiSciascio:2022hyc}.

The design of the CONDOR array unit follows the precedents set by the IceCube  ``IceTop'' upgrade~\cite{ICETOP} and the Pierre Auger Observatory upgrade~\cite{AUGER}. Each unit is enclosed in a 5 mm thick aluminum enclosure with an adjustable-height support. The electronics are based on the design used in the IceCube uDAQ board. The total weight of a unit is less than 40 kg. In a future upgrade, CONDOR may incorporate underground panels to tag muons for veto purposes, employing an approach similar to that of the AMIGA panels in the Pierre Auger Observatory~\cite{Sanchez:2011zza}. 

\begin{figure}[ht!]
\centering
\includegraphics[width=0.5\linewidth]{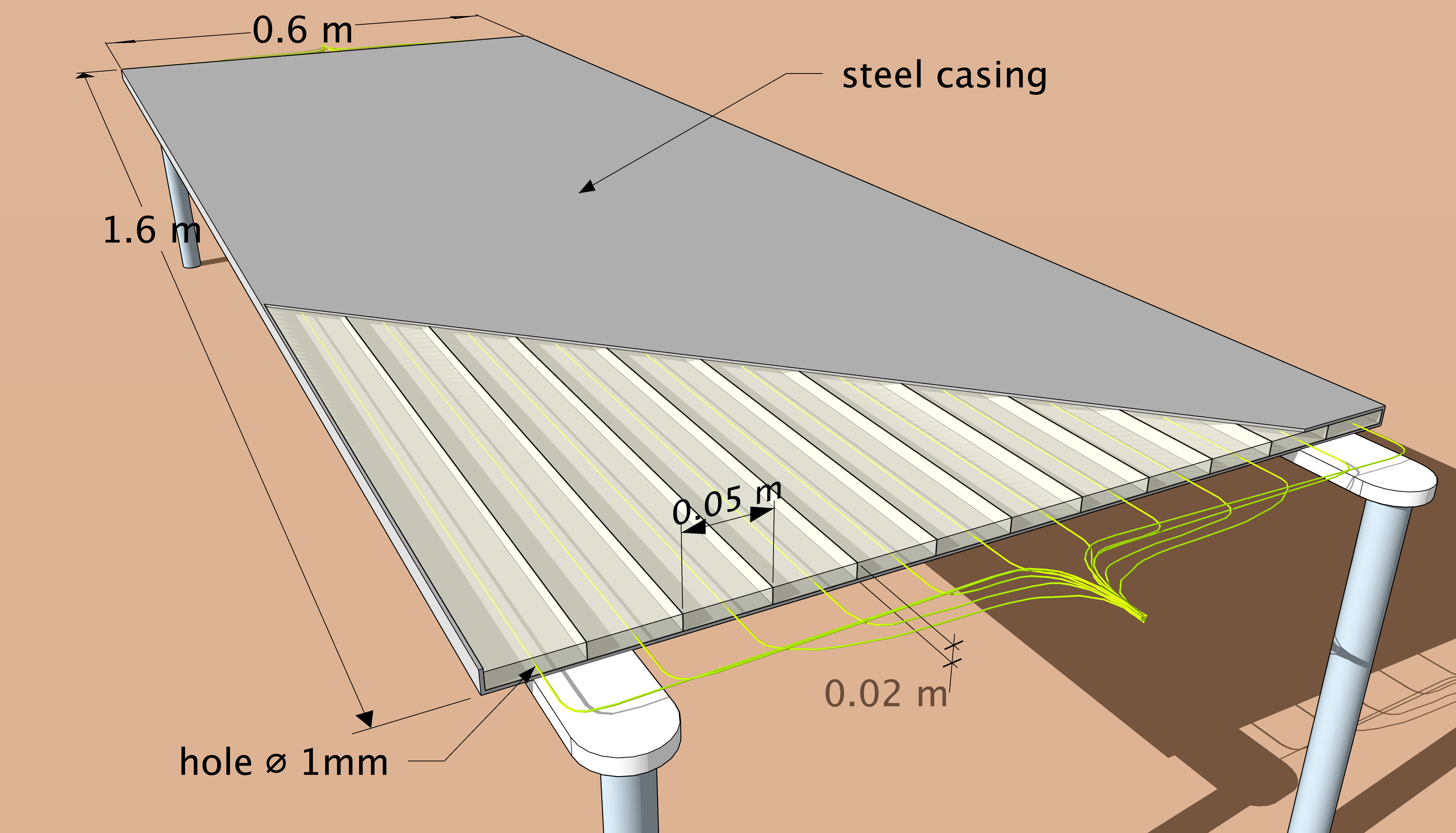}
\caption{CONDOR array unit. }
\label{fig:detailed-group-cutaway}
\end{figure}

The detection area will consist of modules built from scintillator bars modeled after the Pierre Auger SSD, produced at Fermilab National Laboratory using low-cost extruded plastic scintillator ~\cite{Pla-Dalmau:2000puk}. Each bar is coated with a 0.25 mm thick layer of titanium dioxide to enhance light reflection. A central hole with a diameter of 1.0 $\pm$ 0.2 mm runs through each bar, housing a wavelength-shifting fiber air-coupled to the scintillator as shown in Figure~\ref{fig:detailed-group-cutaway}. The fibers from each panel are bundled and coupled to a $6 \times 6$  mm$^{2}$, 15 $\mu$m pitch SiPM (Hamamatsu 14160-6015PS) using optical glue. The Hamamatsu 14160-6015PS SiPM model features 159,565 pixels and a photon detection efficiency (PDE) of 32$\%$, offering significant improvements over the previous generation Hamamatsu S13360-6025PE model used in the IceCube IceTop panels. The older model had a 25 $\mu$m pixel pitch, 57,600 pixels, and a PDE of 25$\%$. Tests with other models in the S14160 and S13360 series, featuring larger microcell sizes (e.g., 25–75 $\mu$m) and higher photon detection efficiencies (up to 50\%), show promising results. The final sensor choice will be guided by optimization studies balancing PDE, saturation behavior, dynamic range, and cost considerations.

A key challenge in having a distributed network of detector arrays extending over a large area is time syncronization, which is necessary for accurate reconstruction of shower direction, triggering, and background rejection. The intrinsic time precision of each CONDOR unit is expected to be less than 
one nanosecond so an accuracy of sub-nanosecond precision and accuracy for the entire array would maximize the performance of the detector array and is set as a requirement. 

White Rabbit (WR) technology~\cite{WhiteRabbit} can be used to synchronize the timing of all units using the IEEE 1588 protocol. WR has been adopted in various experiments, including the IceCube surface array, LHAASO and others. 

A possible layout of the WR network for CONDOR could be as follows. Assuming 6000 CONDOR units, about 600 White Rabbit nodes will be distributed across the CONDOR area through rugged outdoor optical fiber, distributing pulse-per-second (PPS) signals with sub-nanosecond accuracy to groups of about 10 panels. To reduce system complexity and cost, we are investigating the use of a passive fanout board to distribute the PPS signal to each panel readout board via SMA connectors. While this approach offers simplicity, it may introduce timing dispersion, necessitating validation under controlled laboratory conditions. As part of future design iterations, we will also evaluate active fanout solutions and hierarchical switch-based synchronization architectures, guided by system-level performance requirements. A central DAQ hut will host a standard electronic rack that hosts necessary WR switches, which we estimate would be about 35, to synchronize 600 WR nodes. A master WR switch connects to an outdoor GPS antenna, although a rubidium atomic clock could serve as an alternative. The central hut will also host an industrial PC running Linux connected to the internet via Ethernet using the Atacama Astronomical Park network.

The power consumption of the CONDOR observatory is expected to be primarily driven by the WR nodes. Each WR node consumes less than 7 W, and each WR switch consumes less than 80 W. The DAQ industrial computer is expected to require approximately 250 W. Assuming the proposed node-to-detector mapping it yields to a total of approximately 7250 W. This corresponds to $\approx 12$ W per acquisition node or 1.2 W per detector unit on average. Power will be supplied either by solar panels and batteries or through alternative solutions provided by the infrastructure of the Atacama Astronomical Park.

\section{Expedition}

\begin{figure}[ht!]
\centering
\includegraphics[width=0.8\linewidth]{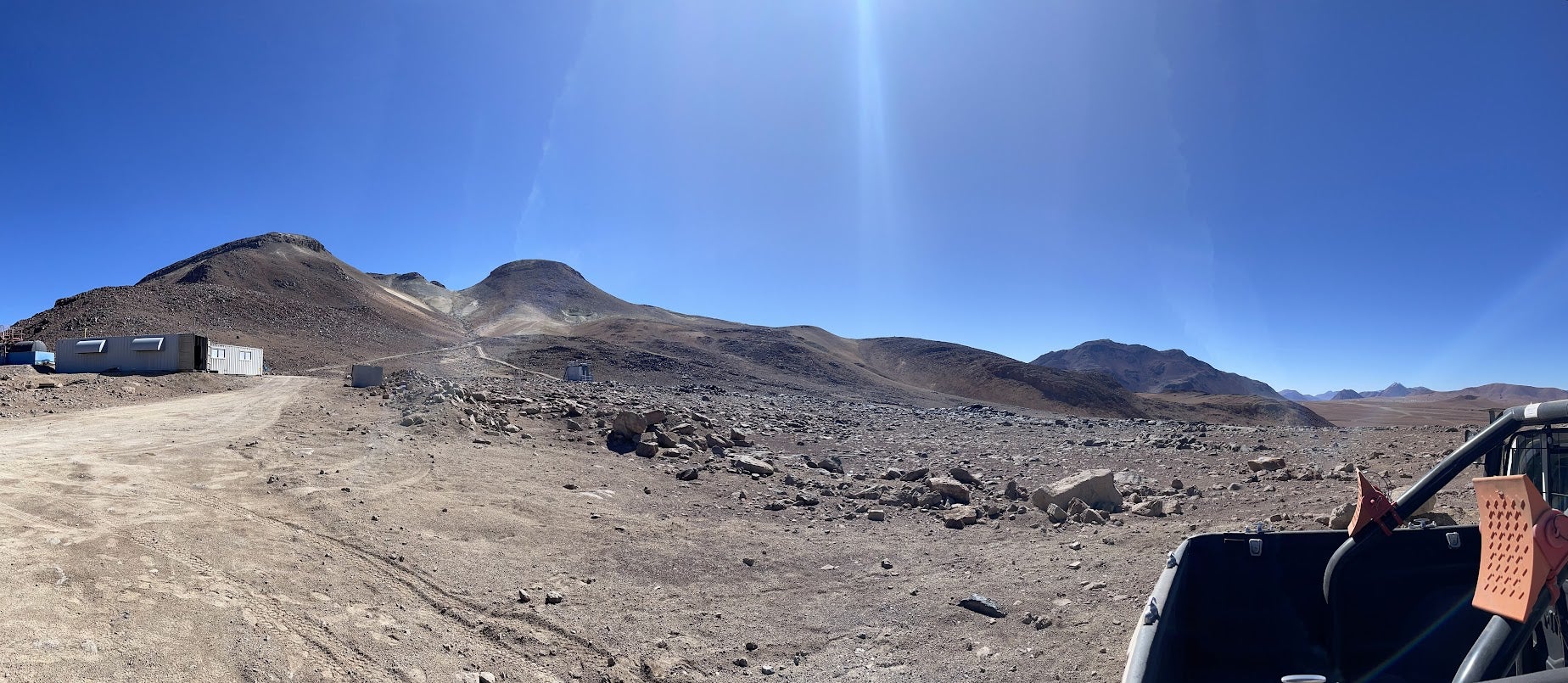}
\caption{Proposed location of the CONDOR array at Cerro Toco, Atacama, Chile. This photograph was taken during the site selection visit in January 2023.}
\label{fig:expedition}
\end{figure}

We conducted an expedition and site-selection visit to the Atacama Astronomical Park in January 2023. After carefully evaluating potential sites based on criteria such as terrain suitability, logistical considerations, and other factors, we selected a location at Cerro Toco ($22^{\circ},57',25'',\text{S};67^{\circ},46',54'',\text{W}$) as the proposed site for the CONDOR observatory (see Figure~\ref{fig:expedition}).

\begin{figure}[ht!]
\centering
\includegraphics[width=0.4\linewidth]{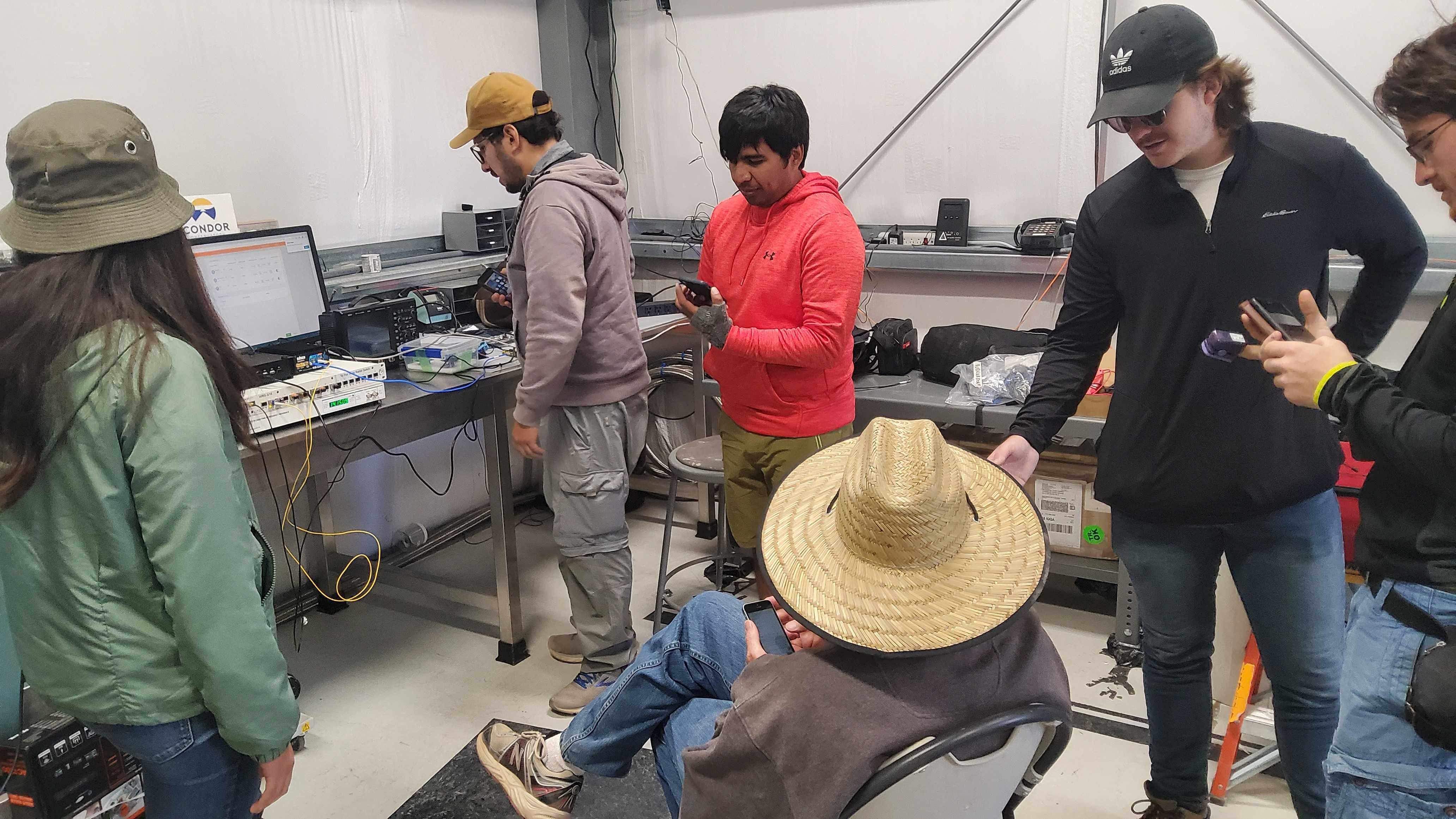}
\includegraphics[width=0.22\linewidth]{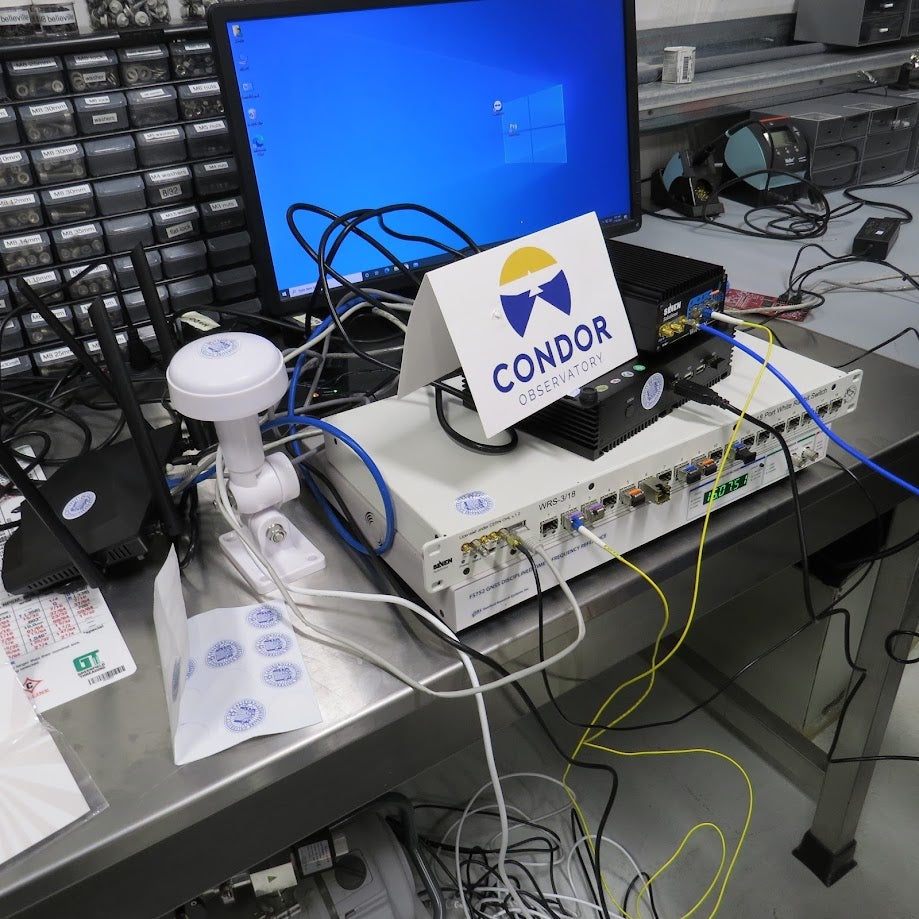}
\includegraphics[width=0.215\linewidth]{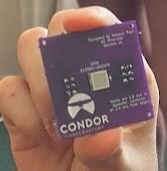}
\caption{CONDOR team (left) deploying a full-chain test setup, including a PC, SiPM with readout electronics, White Rabbit switch, and GPS antenna (middle). A close-up of the SiPM board is shown on the right. These photographs were taken at the CLASS telescope facility on Cerro Toco, Atacama, Chile, in January 2023.}
\label{fig:expedition_setup}
\end{figure}

During this visit, we deployed a complete readout and bias-testing setup consisting of two SiPM boards read out with uDAQ boards, a White Rabbit node and master switch, and a GPS antenna, all remotely operated via a PC (see Figure~\ref{fig:expedition_setup}). Power and workspace were provided by the crew of the CLASS telescope at Cerro Toco, Atacama~\cite{CLASS_2024}. This test aimed to verify the remote operability of all key components and assess the stability of the SiPMs and electronics, noise, and pedestals under realistic atmospheric conditions.

The SiPM was coupled to a small plastic scintillator and enclosed with black tape to ensure light-tightness. An operational voltage of 58~V was applied to the SiPM using the uDAQ board, corresponding to the recommended operating voltage—5~V above its nominal breakdown voltage. The SiPM was placed inside a warehouse without environmental controls, providing representative measurements of ambient conditions at Atacama.

The initial field test was limited to daytime, manual data collection, and was intended to validate basic DAQ operation under outdoor conditions. Environmental data (including temperature) were not yet integrated into the system. Future iterations will include onboard temperature sensors and support continuous, automated data collection, enabling precise temperature-rate correlations and dynamic bias adjustment.

\begin{figure}[h!]
    \centering
    \includegraphics[width=0.38\linewidth]{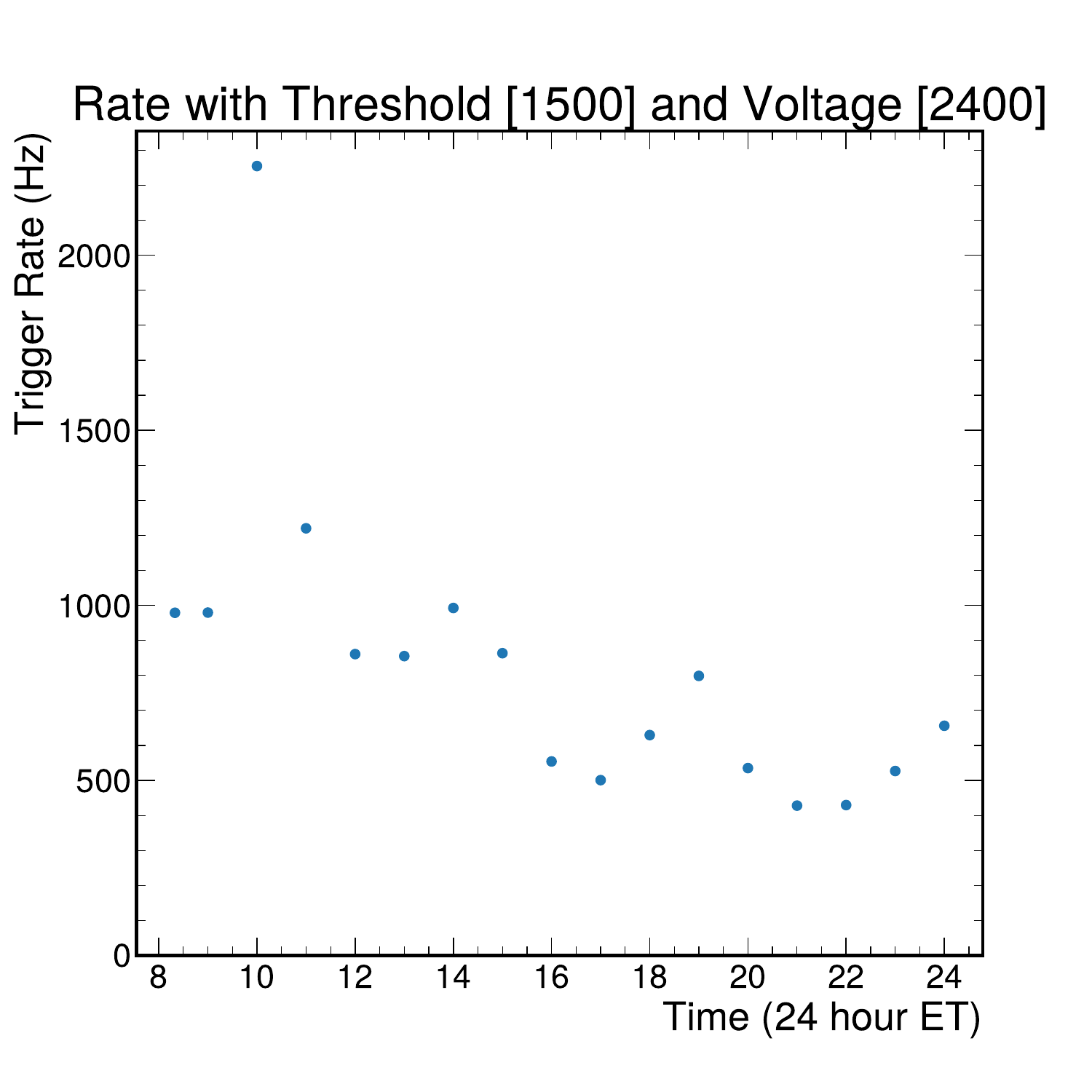}
    \includegraphics[width=0.38\linewidth]{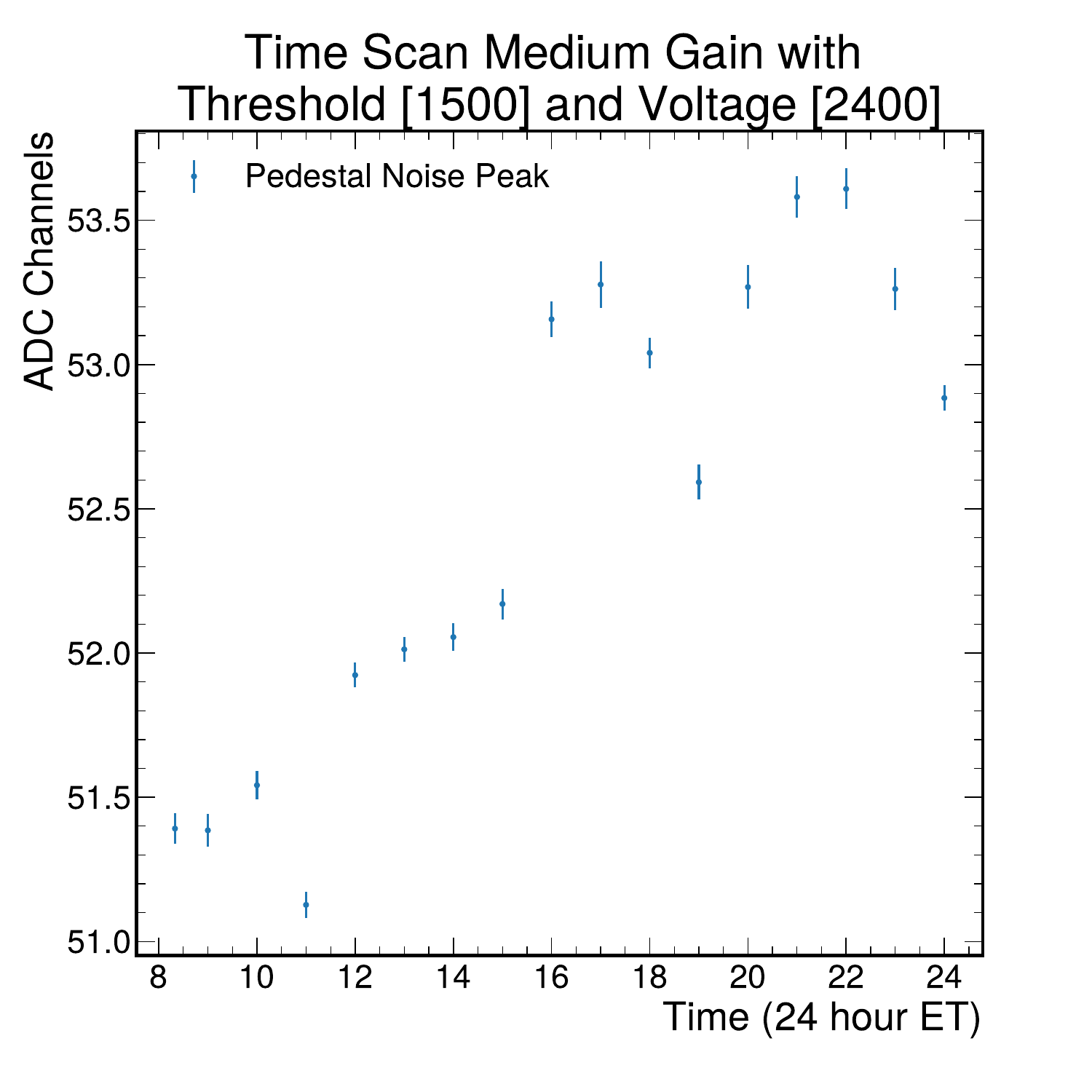}
    \caption{Noise rate (left) and mean pedestal at medium gain (right), both measured in ADC units from the uDAQ board, as a function of time. The voltage applied to the SiPM corresponds to its recommended operational voltage. The threshold was maintained at a fixed ADC value.}
    \label{fig:expedition-measurements}
\end{figure}

Figure~\ref{fig:expedition-measurements} shows example measurements taken during a representative test day, highlighting significant changes in the trigger rate induced by substantial temperature variations typical in Atacama. These results indicate that achieving a stable noise level requires implementing a temperature-dependent bias voltage scheme. Such a scheme could be operated remotely using feedback from local temperature sensors, which we plan to incorporate into the setup.

Additionally, we are developing a system to integrate LED-based calibration sources in each detector module. This will enable periodic in-situ characterization of SiPM gain, dark noise, and crosstalk, similar to the “sensor finger plot” approach described in~\cite{Heller:2016rlc}. These features are essential for optimizing the photon detection performance and ensuring long-term stability of the detector modules.

Overall, the CONDOR site-selection expedition and chain test successfully completed its mission. This expedition marked an important first step toward demonstrating the remote operation of key components of CONDOR under realistic conditions and finalizing the selection of the CONDOR observatory site. Additional results will be presented in a separate publication.

\FloatBarrier
\section{Simulated performance of the CONDOR observatory}
\label{sec1}

\subsection{Monte Carlo Generator}
The CORSIKA (COsmic Ray SImulations for KAscade) \cite{CORSIKA} software, a widely used Monte Carlo framework, has been employed to simulate EAS for the CONDOR Observatory. This simulation framework is integral to studying the interactions of CR with the atmosphere and optimizing the design and performance of the detector array, the observatory's unique environmental conditions—high altitude, low atmospheric pressure, and the local geomagnetic field—have been incorporated into the CORSIKA simulations to ensure accuracy and realism.

The simulations focus on two primary CR types: gamma rays and protons. These were chosen to reflect the key scientific goals of the observatory, including gamma-ray astronomy and CR composition studies. Showers were simulated across a range of energies, from $20$ GeV to $800$ GeV, and zenith angles ranging from $0^\circ$ to $60^\circ$ in steps of $2^\circ$. At each angle and specific energy of the CR, $10^{3}$ showers were simulated with different seeds.

In our study at the CONDOR Observatory, we employ the Energy-conserving Quantum Mechanical multiple-Scattering approach (EPOS) model~\cite{PierogWerner2008} as the high-energy hadronic interaction model for simulating CR interactions within the energy range of $100$ GeV to $1$ TeV. EPOS is selected for its comprehensive approach to particle physics, adeptly modeling both the hard processes of high-energy collisions, such as minijet production, and the subsequent soft interactions, including string fragmentation and resonance decays. This model uniquely integrates the physics of partons, off-shell remnants, and nuclear effects like shadowing, thereby providing a detailed simulation from the initial interaction to the complete development of the air shower. Its accuracy within this specific energy spectrum is well-established, closely matching experimental data \cite{PierogWerner2008, Pierog2013}. The model's capability to conserve energy while tracking particle production and decay ensures that the simulated showers accurately reflect real-world observations, especially in terms of muon production and the longitudinal structure of showers.

The GEISHA~\cite{GEISHA} model is used for low-energy interactions. GEISHA is particularly effective for modeling electromagnetic and low-energy hadronic processes, where precision in the treatment of low-energy secondary particle interactions is essential. This model is specifically designed to handle low-energy phenomena, including particle production, decay, and scattering, which are critical for accurately simulating the early stages of air showers ~\cite{Fesefeldt1985, Geant4}. Its use complements EPOS in providing a complete simulation framework spanning both low- and high-energy regimes.

Key outputs from the simulations include the lateral and longitudinal distributions of secondary particles, their arrival times at the detector plane, and the energy deposits across the scintillator array. The detector geometry, comprising thousands of plastic scintillators optimized for the detection of charged particles, was integrated into the simulation framework. These detailed outputs are 
used to refine angular reconstruction algorithms and develop $\gamma$-ray/proton discrimination techniques.

Furthermore, the high-altitude location reduces the influence of atmospheric attenuation on the secondary particles, allowing for more of these particles per square meter and enabling precise measurements of particle distributions arriving at the detectors of the CONDOR Observatory. This detailed simulation framework ensures that the specific environmental and detector conditions of the observatory are faithfully represented, providing critical insights into the physics of CR interactions and the performance of the observatory's detection systems.

\begin{figure}[ht!]
    \centering
    \begin{subfigure}{0.495\linewidth}
        \centering
        \includegraphics[width=\linewidth]{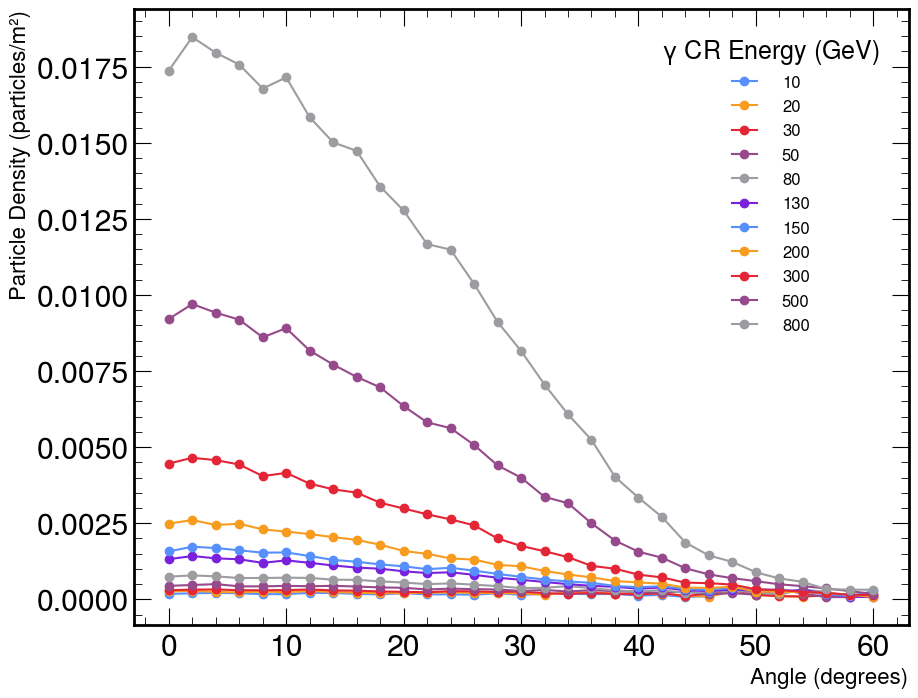}
        \caption{}
        \label{fig:subfig3}
    \end{subfigure}
    \hfill
    \begin{subfigure}{0.495\linewidth}
        \centering
        \includegraphics[width=\linewidth]{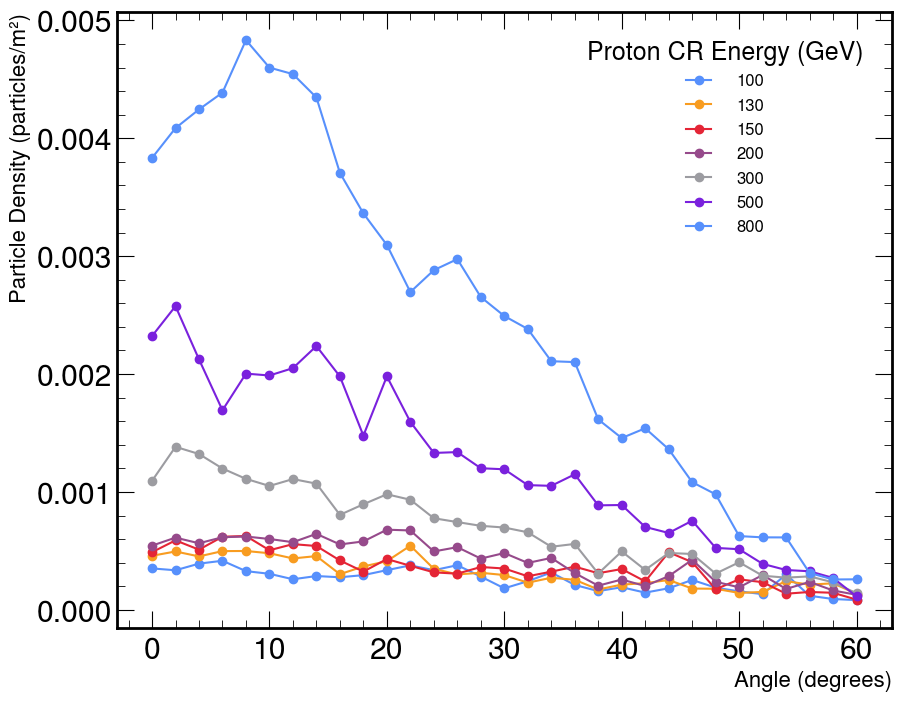}
        \caption{}
        \label{fig:subfig4}
    \end{subfigure}
\caption{(a) Average particle density for the CONDOR Observatory, $\gamma$,(b) Average particle density for the CONDOR Observatory, Protons.}
\label{fig:particle_density}
\end{figure}

Figure~\ref{fig:particle_density} illustrates the average particle count density per square meter at the CONDOR Observatory for CR induced showers simulated with photons and protons as CR across a range of energies and zenith angles. The data highlights a distinct trend: as the zenith angle increases, the particle density decreases. This behavior can be attributed to the increased atmospheric path length at higher angles, which attenuates the secondary particle flux reaching the ground and the smaller effective area of detector that is in the path of the particles.

A clear distinction is observed between photon-induced showers and proton-induced showers. Gamma rays generally produce a higher particle density at the same energy and zenith angle compared to protons. This is because photon-induced showers are more compact, leading to higher localized densities. In contrast, proton-induced showers involve more complex interactions and broader distributions of secondary particles, resulting in lower particle densities. 


\subsection{Zenith angle and Energy reconstruction}

We utilized the simulated data described in the previous section to reconstruct the zenith angle of the primary CR. Our approach involves fitting the shower front's planar structure in the \((x, y, t)\) coordinate system, capturing the arrival times and positions of secondary particles detected across the array. This methodology allows us to determine the best-fit plane representing the shower front and infer the zenith angle. While conical shower front fits were also explored, they did not yield significant improvements in angular resolution at the low energies considered, due to the limited number of detected particles. As a result, planar fitting was adopted as the baseline reconstruction method.

The first step in this process involves calculating the centroid of the shower particles arriving at the detectors in the \((x, y)\) plane. The centroid is determined as the average position of all detected particles weighted by their respective energy deposits, ensuring that the central region of the shower core, where the particle density is highest, is accurately represented. This centroid provides the central reference point for subsequent planar fits, providing a solid basis for reconstructing the shower geometry. 

\begin{figure}[ht!]
    \centering
    \begin{subfigure}{0.495\linewidth}
        \centering
        \includegraphics[width=\linewidth]{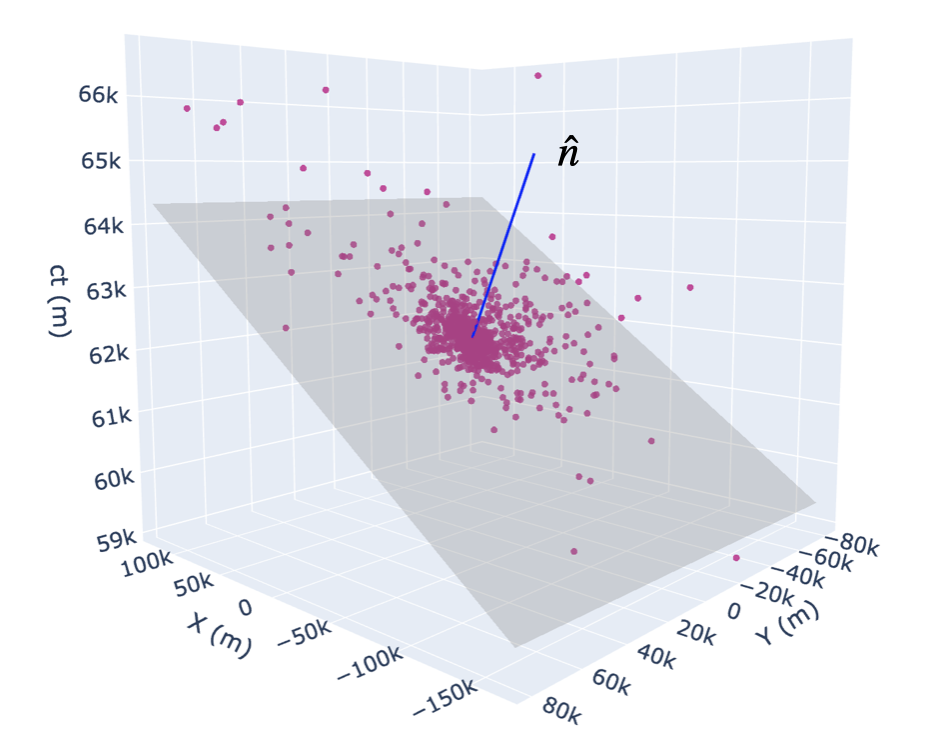}
        \caption{}
        \label{fig:subfig5}
    \end{subfigure}
    \hfill
    \begin{subfigure}{0.495\linewidth}
        \centering
        \includegraphics[width=\linewidth]{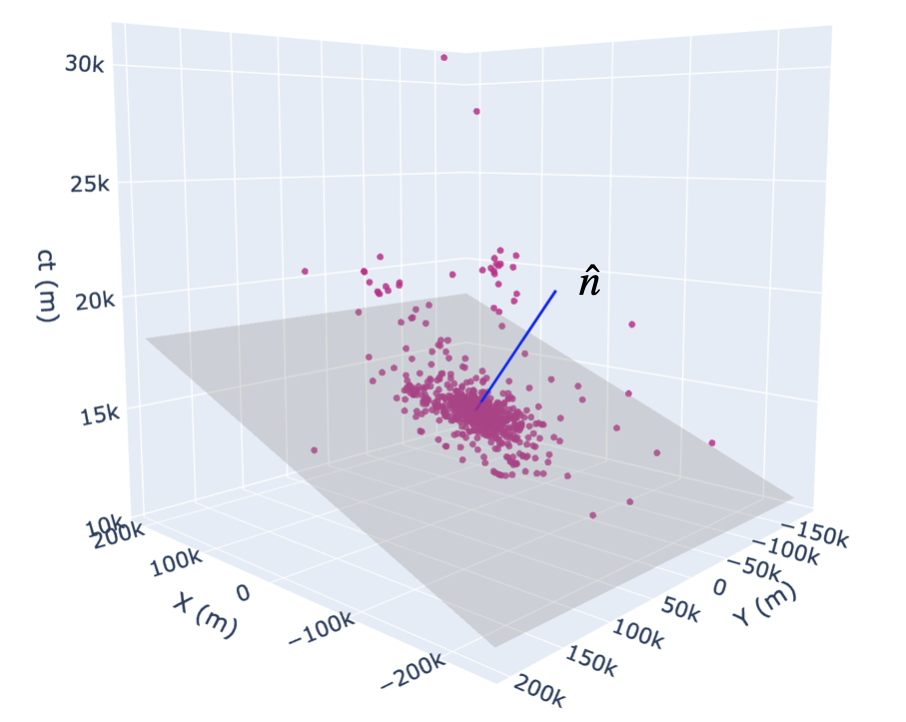}
        \caption{}
        \label{fig:subfig6}
    \end{subfigure}

\caption{Shower distribution in the coordinate system used for the simulation.  
    (a) Photon, (b) Proton. Both with $150$ GeV with zenith angle equal to 45$^{\circ}$. A planar fit (in grey) to the shower plane and its normal vector in blue line are shown.}
\label{fig:angle_reconstruction0}
\end{figure}

Planar fits are performed across the entire detector array as well as sub-regions of it. As example,  Figure~\ref{fig:angle_reconstruction0} shows the particle distribution at the Observatory level for gamma and proton for a specific CR energy and zenith angle. This subdivision is crucial for capturing the planar behavior of the shower front across varying scales, allowing us to account for potential deviations from a purely planar structure due to geomagnetic effects, detector resolution, or atmospheric scattering. The fit minimizes the deviations of the detected particle arrival times from a plane in the \((x, y, ct)\) coordinate system, where $c$ is the speed of light in $[m/ns]$ and $t$ is the arrival times (in $[ns]$) of the particles to the detectors coordinate system, effectively modeling the temporal evolution of the shower front. 

The normal vector to the fitted plane was calculated relative to the centroid of the particle distribution, representing the central ray of the hadronic cascade. The zenith angle corresponds to the angle formed between the normal of the shower and the \(z\)-axis, calculated using the following formula:
\[
\theta = \arccos\left(\frac{\vec{n} \cdot \hat{z}}{|\vec{n}| |\hat{z}|}\right)
\]
where \(\vec{n}\) is the normal vector to the fitted plane and \(\hat{z}\) is the unit vector in the \(z\)-direction. This calculation allows us to determine the zenith angle between the vertical axis and the central ray of the cascade.

For each combination of zenith angle and primary particle energy a large number of fits are calculated. By considering multiple fits for each simulation, we ensure robustness in the reconstruction process and mitigate the impact of statistical fluctuations or outliers in the simulated data. 
With this large number of dataset we perform different cuts in the $x$ $y$ coordinates. The first cut included the entire detector array, the second considered only the central detector region, and subsequent cuts focused progressively on smaller regions moving inward to the detector center. 
These cuts end in a region very near the shower centroid, which approximates a flat surface very well. Among all these cuts, the zenith angle corresponds to the fit with the smallest $\chi ^2$ of the planar fits, ensuring the most accurate representation of the shower geometry. This selection process ensures that the reconstructed zenith angle closely represents the actual inclination of the primary CR. 

This iterative procedure provides a detailed reconstruction of the zenith angle for a wide range of CR energies and arrival directions. The methodology is particularly effective in high-altitude environments, such as the CONDOR Observatory, where the reduced atmospheric attenuation and higher particle densities improve the accuracy of the zenith angle reconstruction. At these altitudes, the thinner atmosphere minimizes scattering and energy loss of the secondary particles, leading to sharper and more distinct shower fronts that are ideal for planar fitting techniques.  

By leveraging these simulations and applying planar fitting methods, we achieve a high-precision determination of the zenith angle. This is critical for studying the origins, arrival directions and properties of CR.

\begin{figure}[ht!]
    \centering
    \begin{subfigure}{0.495\linewidth}
        \centering
        \includegraphics[width=\linewidth]{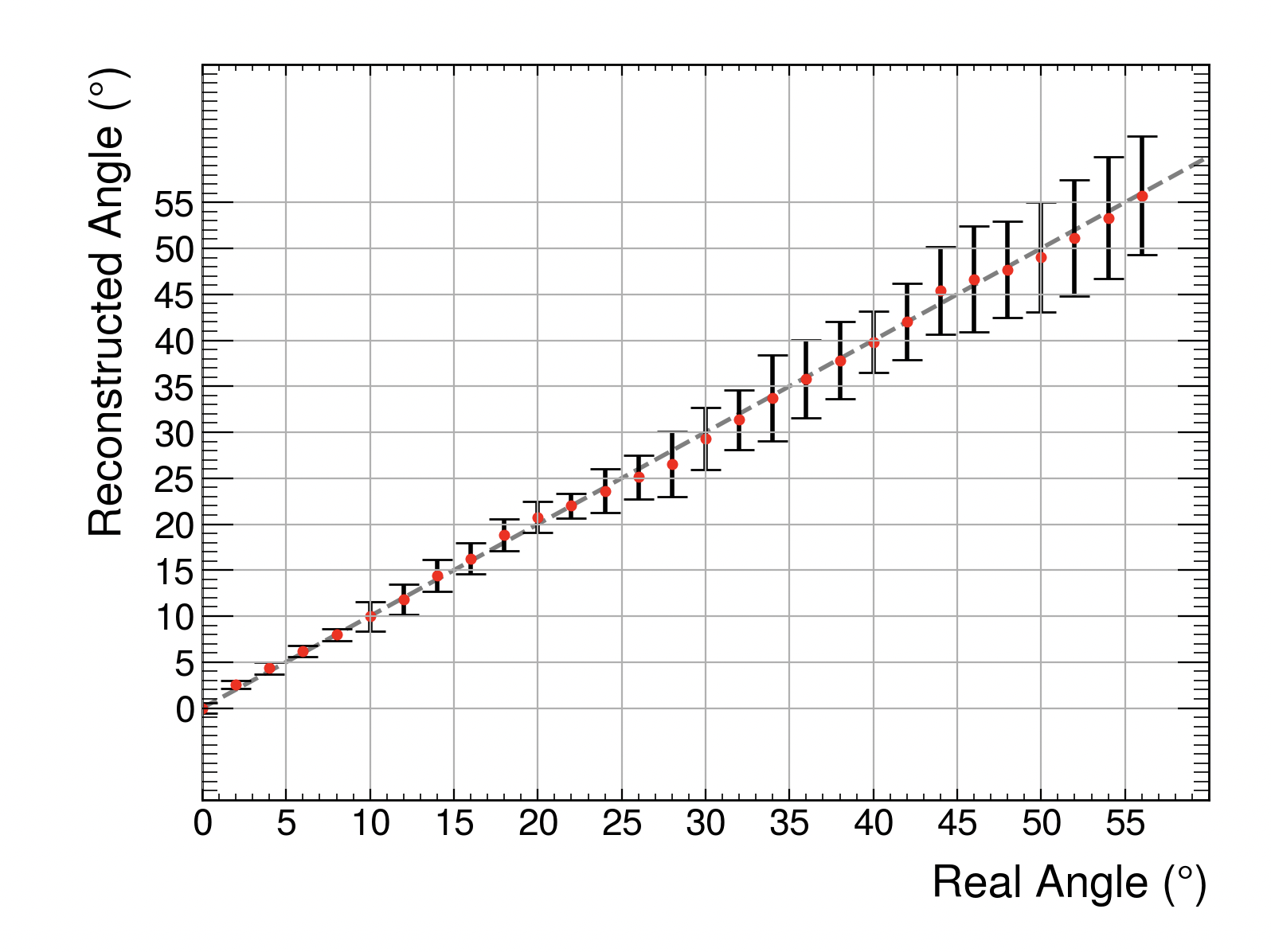}
        \caption{}
        \label{fig:subfig7}
    \end{subfigure}
    \hfill
    \begin{subfigure}{0.495\linewidth}
        \centering
        \includegraphics[width=\linewidth]{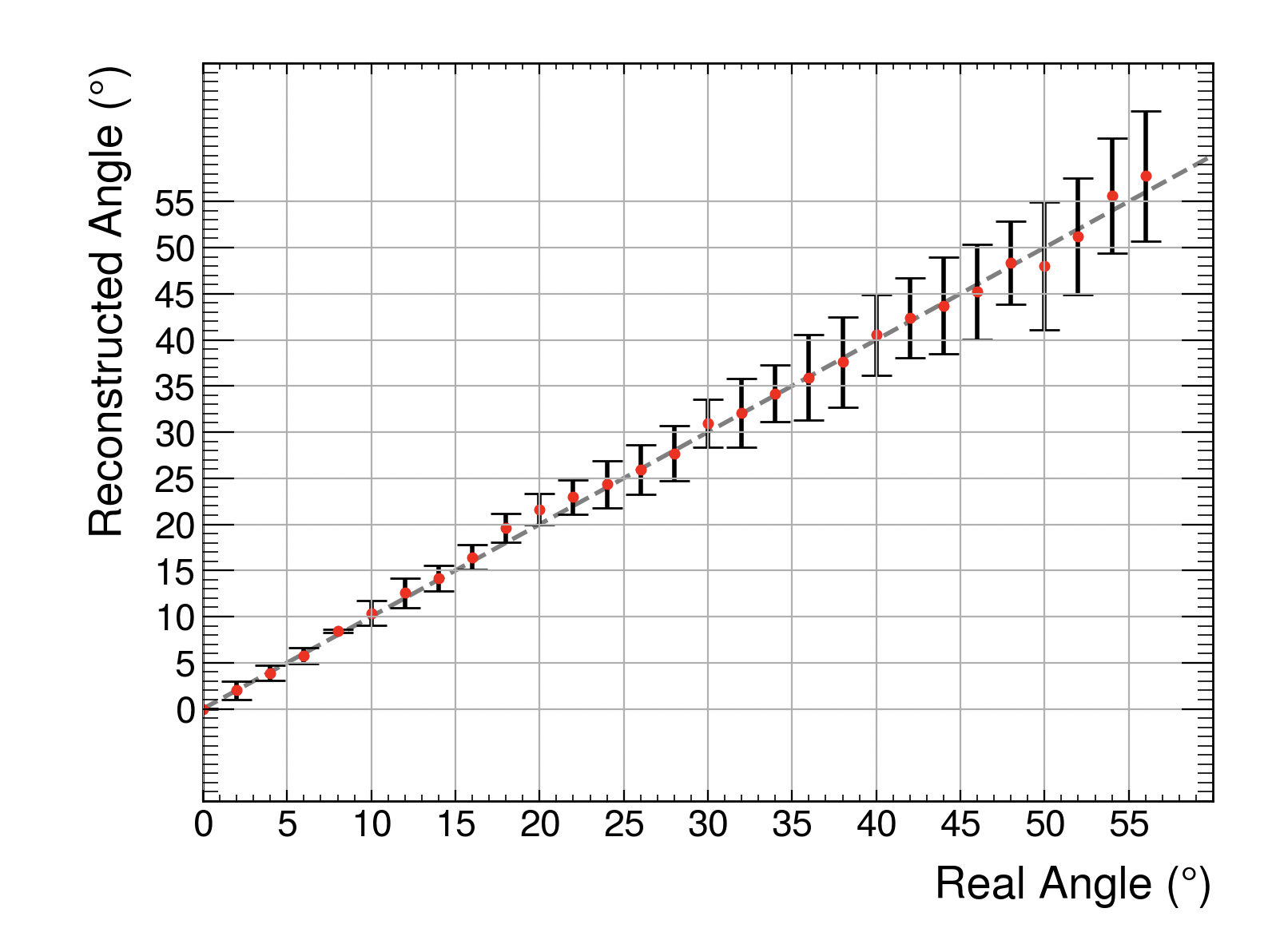}
        \caption{}
        \label{fig:subfig8}
    \end{subfigure}

\caption{Comparison between reconstructed angles and true angles: (a) 150 GeV Photon (b) 150 GeV Proton.}
\label{fig:angle_reconstruction}
\end{figure}

Figure~\ref{fig:angle_reconstruction} shows the relationship between the true zenith angles and the reconstructed angles, revealing an approximately linear correlation for photon-initiated and proton-initiated CR showers. 
Reconstructed angles were analyzed in steps of 2$^{\circ}$ from 0$^{\circ}$ to 60$^{\circ}$, showing that while the predicted angles closely follow the true values, the reconstruction errors increase with larger zenith angles. This trend is primarily due to the reduced number of particles reaching the detector as the zenith angle increases. At larger angles, the shower traverses a greater atmospheric depth, causing significant attenuation of the secondary particle flux.

Despite the growing uncertainties, the angular predictions remain reasonably accurate, albeit with larger errors resulting from the limited particle counts. Notably, the reconstruction process was feasible up to 56$^{\circ}$ for photon and proton showers. Beyond this angle, the number of particles reaching the detector becomes insufficient to perform a reliable fit, making accurate angle reconstruction impractical. These findings highlight the challenges of reconstructing high zenith angle showers and emphasize the importance of particle statistics in achieving precise angular determinations.

This analysis was conducted for photon-initiated and proton-initiated showers, covering an energy range of 20 to 800 GeV for photons and 150 to 800 GeV for protons.

The reconstructed energy values were obtained using a deep learning (DL) model designed to analyze data from simulated Extensive Air Showers (EAS) under the CONDOR Observatory conditions. Specifically, a hybrid architecture was employed that combines convolutional neural networks (CNNs) and transformers. This architecture merges the strengths of both models to overcome their individual limitations. CNNs are highly effective at extracting local features, such as the relationships between nearby detector hits, while transformers utilize a self-attention mechanism to capture global, long-range dependencies across the entire data sequence~\cite{10.1093/pasj/psad071,11129604}.
The model processes the complete spatiotemporal sequence of detector activations, including the arrival time, deposited energy, and position of each hit, to extract complex patterns. This integrated approach leverages both the local and global information contained within the shower cascade.  

\begin{figure}[ht!]
    \centering
    \includegraphics[scale=0.5]{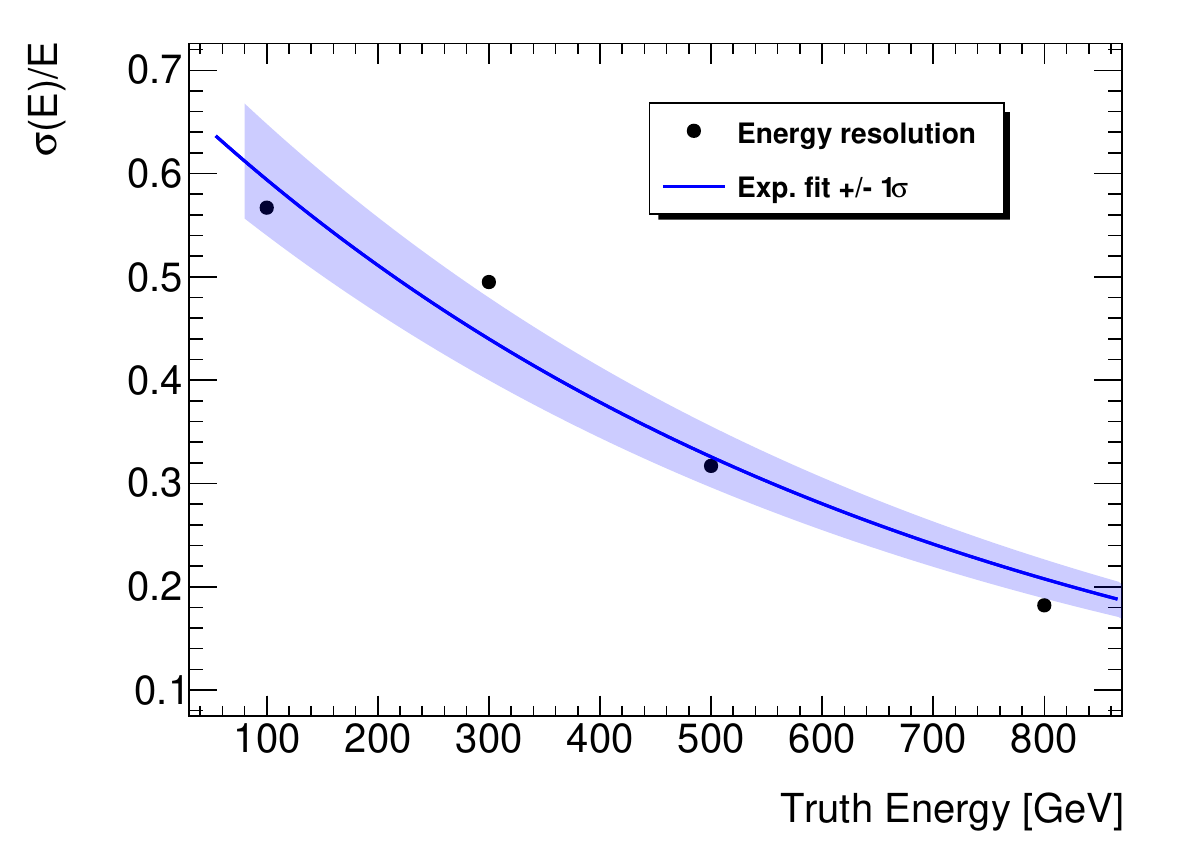}
    \caption{\small{
    Energy reconstruction resolution as a function of true energy for four fixed incident energies, as obtained using a machine learning (ML) algorithm. The points are fitted with an exponential function to guide the trend. The resolution improves with increasing energy, as expected due to higher particle statistics and reduced relative fluctuations at larger energies.}}
    \label{fig:E_reso}
\end{figure}

The resolution of the energy reconstruction can be seen in Figure~\ref{fig:E_reso}, where $\sigma(E)$ is the standard deviation of the reconstructed energy distribution at a given truth energy.

\begin{figure}[ht!]
    \centering
    \begin{subfigure}{0.495\linewidth}
        \centering
        \includegraphics[width=\linewidth]{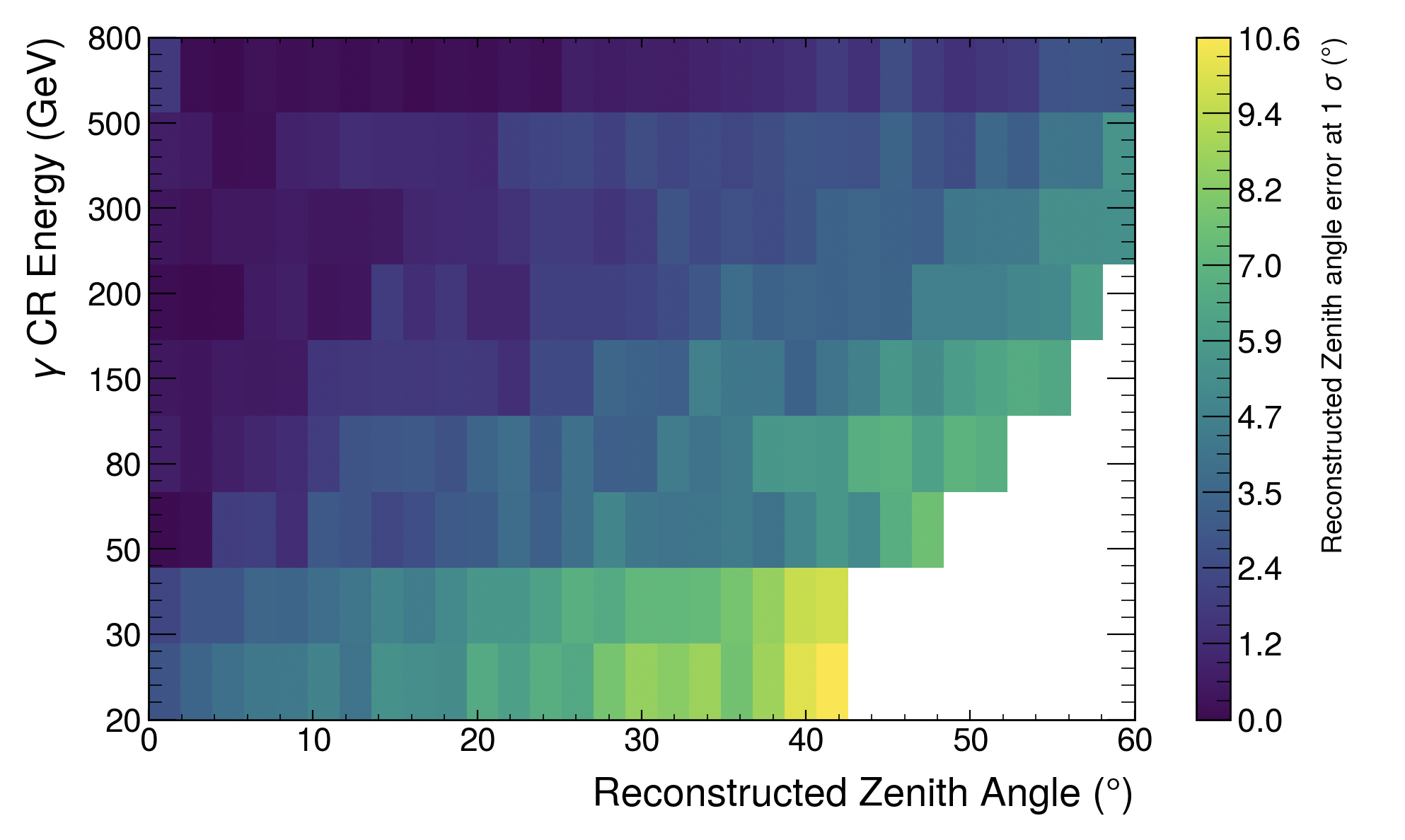}
        \caption{}
        \label{fig:subfig9}
    \end{subfigure}
    \hfill
    \begin{subfigure}{0.495\linewidth}
        \centering
        \includegraphics[width=\linewidth]{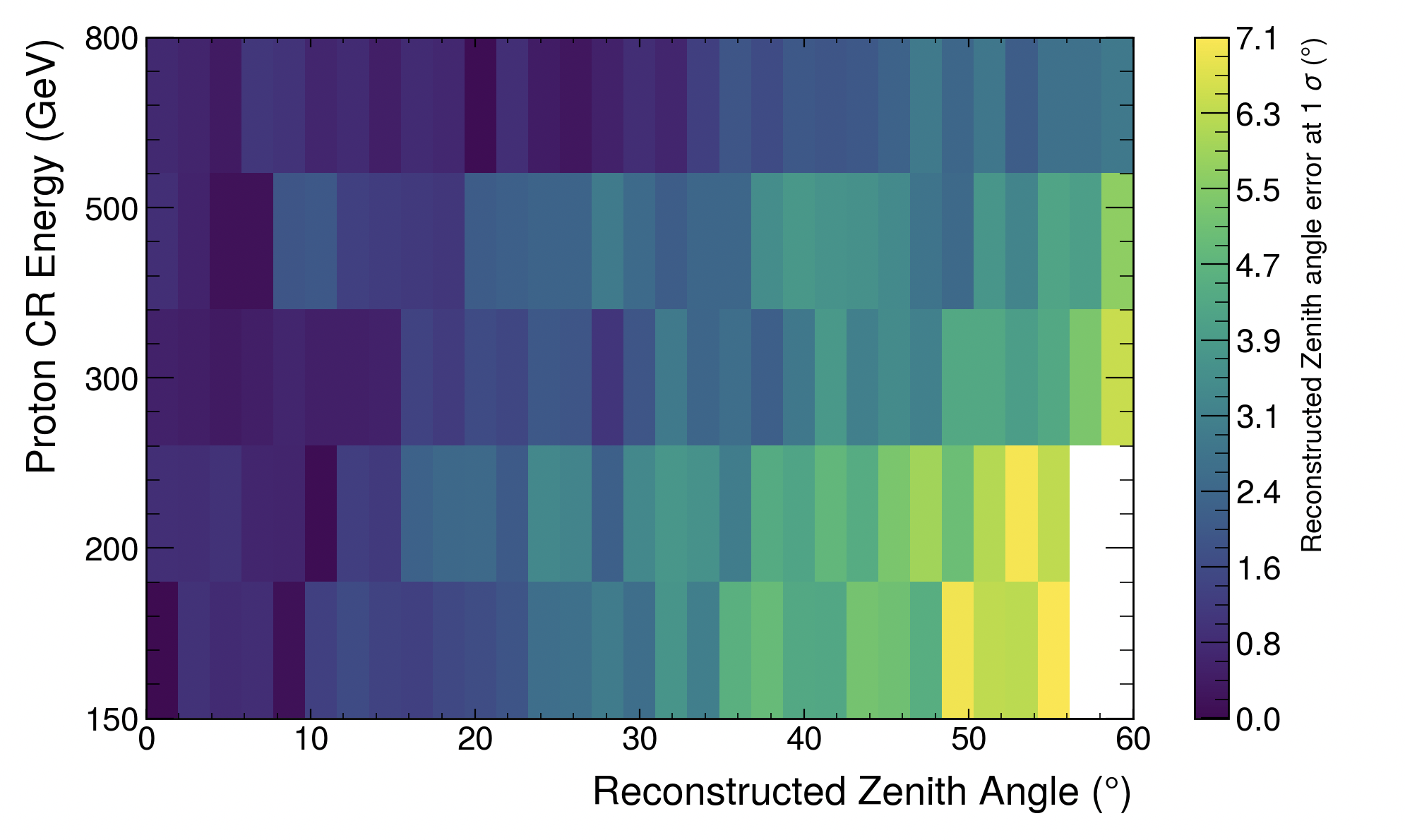}
        \caption{}
        \label{fig:subfig10}
    \end{subfigure}
\caption{Heatmap of angular reconstruction errors as a function of true zenith angle and primary particle energy: (a) Photon (b) Proton.}
\label{fig:particle_errors}
\end{figure}

As observed in Figure~\ref{fig:particle_errors}, the color scale represents the magnitude of the reconstructed error. Photon showers at lower energies exhibit larger errors in the reconstructed angles. Additionally, the range of reconstructed angles is more limited at these low energies. This behavior arises from the reduced number of particles in the shower, making the reconstruction process significantly more challenging.

The reconstruction becomes more accurate at higher photon energies, with smaller errors and a broader coverage of reconstructed angles. A similar trend is evident for proton showers, where higher energy simulations yield improved angular precision and a wider angular reconstruction range. These results highlight the critical role of primary energy in the accuracy and reliability of zenith angle reconstruction for both photon and protons.

The described reconstruction technique builds upon established methods for shower front modeling and particle trajectory reconstruction \cite{Gaisser1990, Knapp2005}. The robustness of the planar fitting approach ensures that the methodology is adaptable to varying detector configurations and environmental conditions, making it an invaluable tool for analyzing CR interactions in the atmosphere.

\subsection{\texorpdfstring{$\gamma-$}-Proton discrimination for induced air shower}
The tagging system developed for the CONDOR Observatory is a crucial tool for classifying CR induced air showers as either proton- or photon-induced. Drawing inspiration from tagging techniques employed in the ATLAS detector at CERN~\cite{ATLAS_Tagging}, this system utilizes Monte Carlo simulations conducted with CORSIKA. 

The algorithm achieves reliable and robust classification by generating high-statistics templates for proton- and photon-induced showers. This capability significantly enhances the observatory’s ability to distinguish between primary CR types, providing deeper insights into CR composition and advancing the study of related astrophysical phenomena.

\subsubsection{Radial and time shower distribution}
\begin{figure}[ht!]
    \centering
    \begin{subfigure}{0.495\linewidth}
        \centering
        \includegraphics[width=\linewidth]{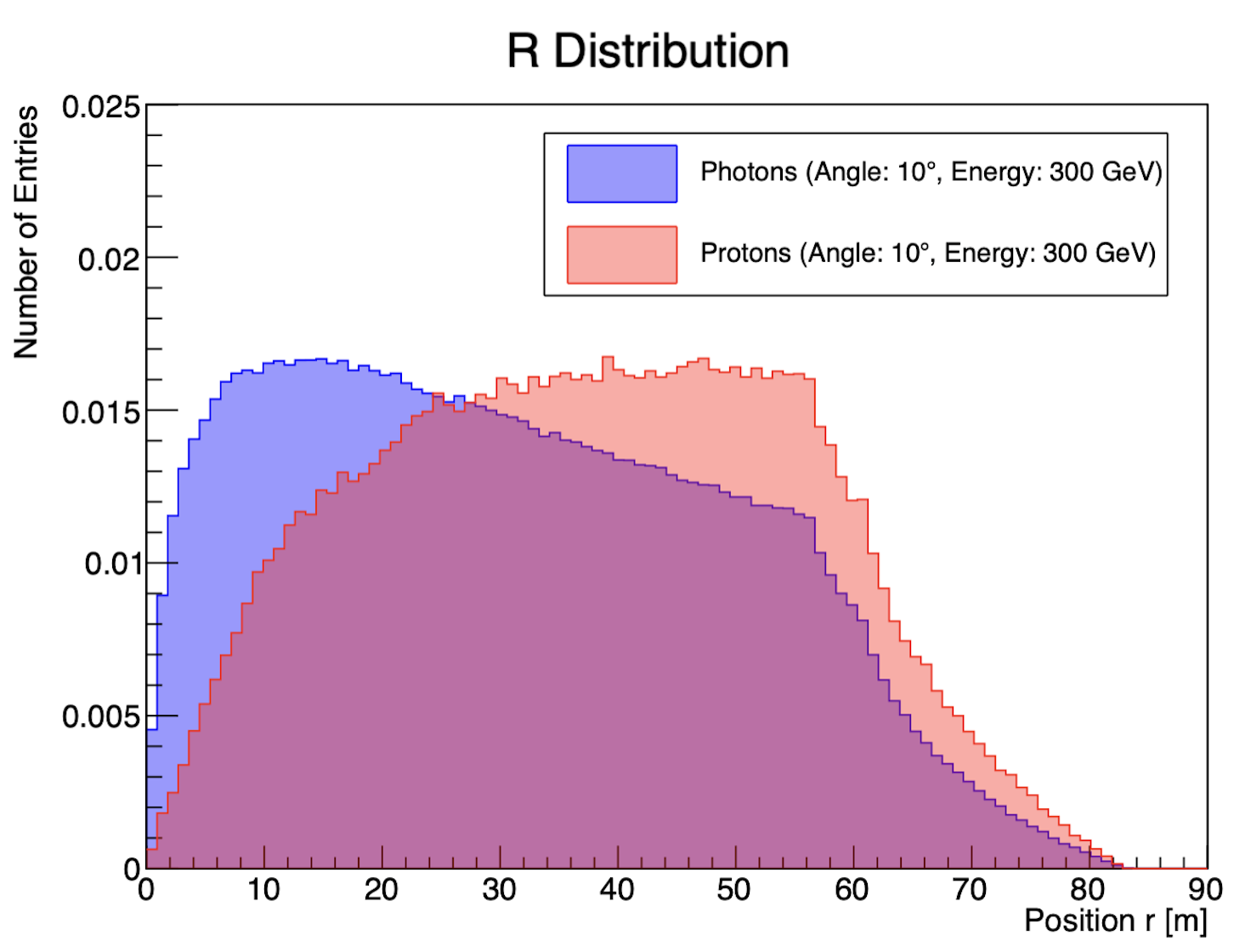}
        \caption{}
        \label{fig:subfig11}
    \end{subfigure}
    \hfill
    \begin{subfigure}{0.495\linewidth}
        \centering
        \includegraphics[width=\linewidth]{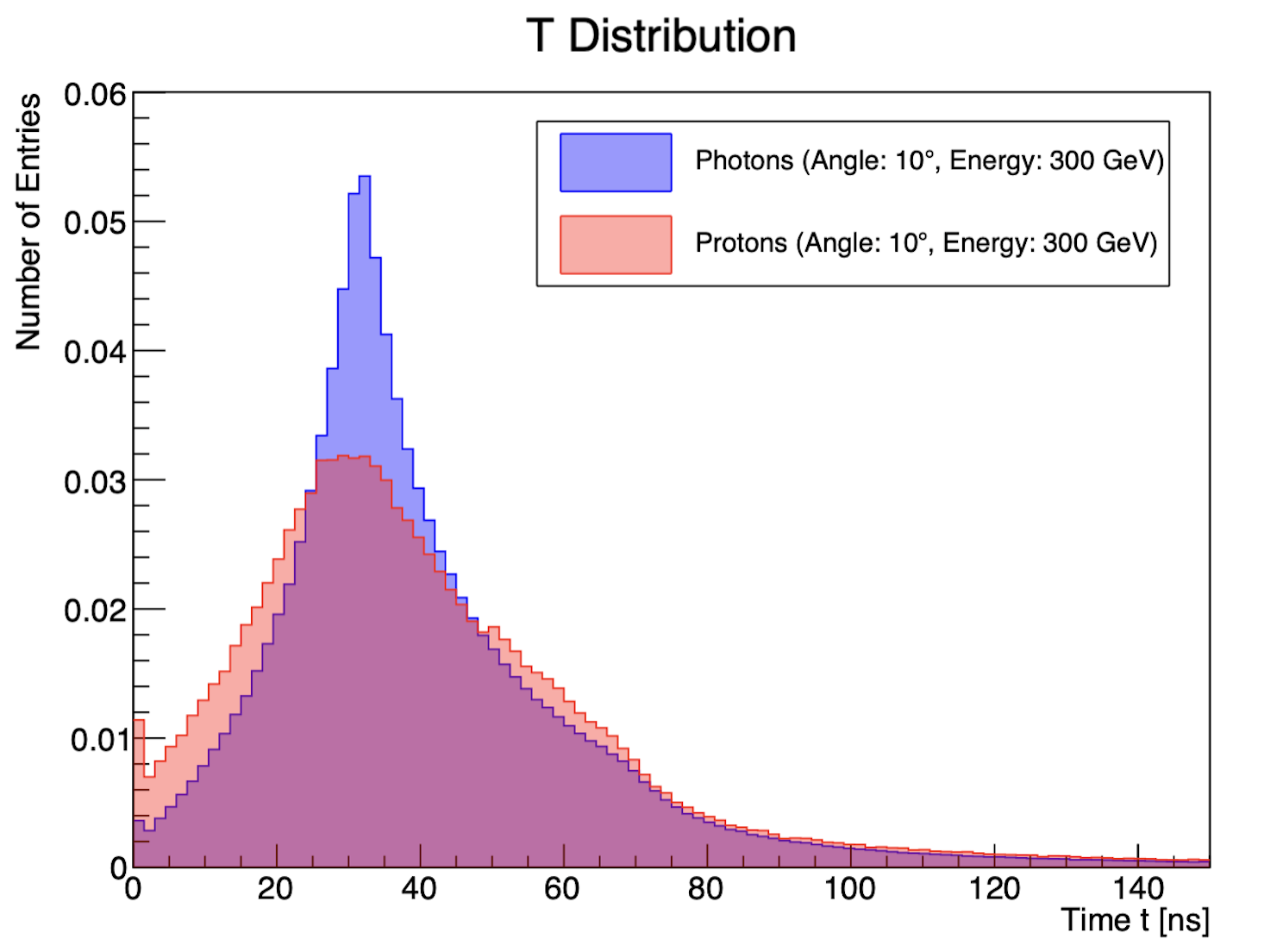}
        \caption{}
        \label{fig:subfig12}
    \end{subfigure}
    \caption{(a) Radial distribution of secondary particles, where the blue histogram represents the statistical template derived from multiple photon showers, while the red histogram corresponds to the template for proton showers, (b) Time distribution of secondary particles, similarly distinguishing photon and proton showers. These templates are used in the statistical fit to classify unknown showers.}
\label{fig:shower_distributions}
\end{figure}

Key observables included the spatial parameter $r = \sqrt{x^2 + y^2}$ and the temporal parameter \(t\), corresponding to the time of particle arrival. High-statistics Monte Carlo templates for proton- and photon-induced showers were constructed by aggregating multiple simulated showers, as shown in Figure~\ref{fig:shower_distributions}.

\subsubsection{Algorithm Overview}
The tagging process involves fitting the observed shower distributions to predefined templates, assigning statistical weights to the proton and photon components. These weights determine the classification outcome, as expressed by:

\[
W_p D_p + W_\gamma D_\gamma = D_C, \quad W_p + W_\gamma = 1,
\]
where \(D_p\) and \(D_\gamma\) represent the proton and photon templates, respectively, and \(D_C\) is the combined distribution. The tagging procedure was evaluated across all simulated energies and incidence angles. 
Among the observables, the spatial parameter \(r\) proved the most effective classification, as the temporal parameter \(t\) exhibited significant overlap between proton and photon distributions.

\begin{figure}[ht!]
    \centering
    \begin{subfigure}{0.495\linewidth}
        \centering
        \includegraphics[width=\linewidth]{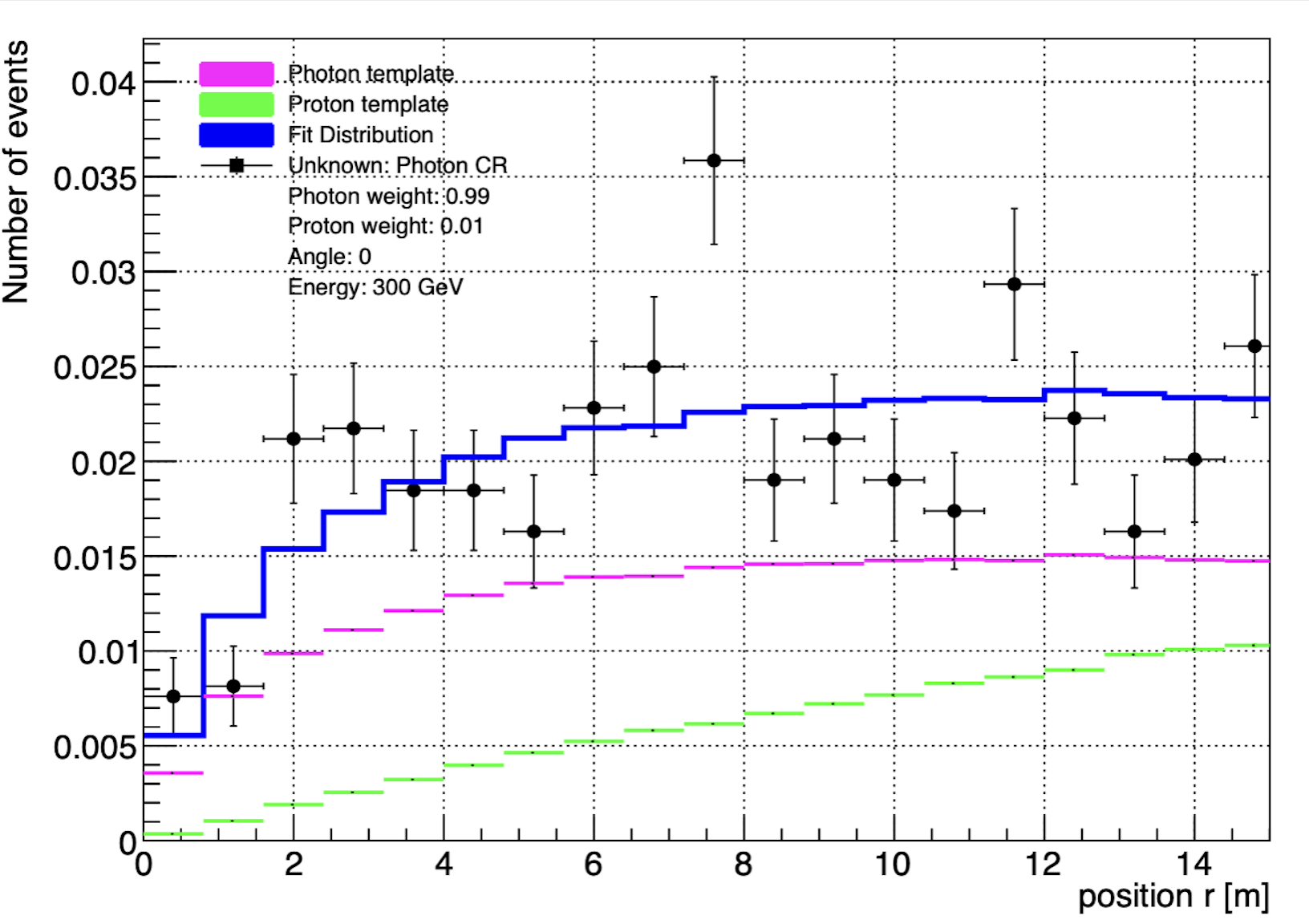}
        \caption{}
        \label{fig:subfig13}
    \end{subfigure}
    \hfill
    \begin{subfigure}{0.495\linewidth}
        \centering
        \includegraphics[width=\linewidth]{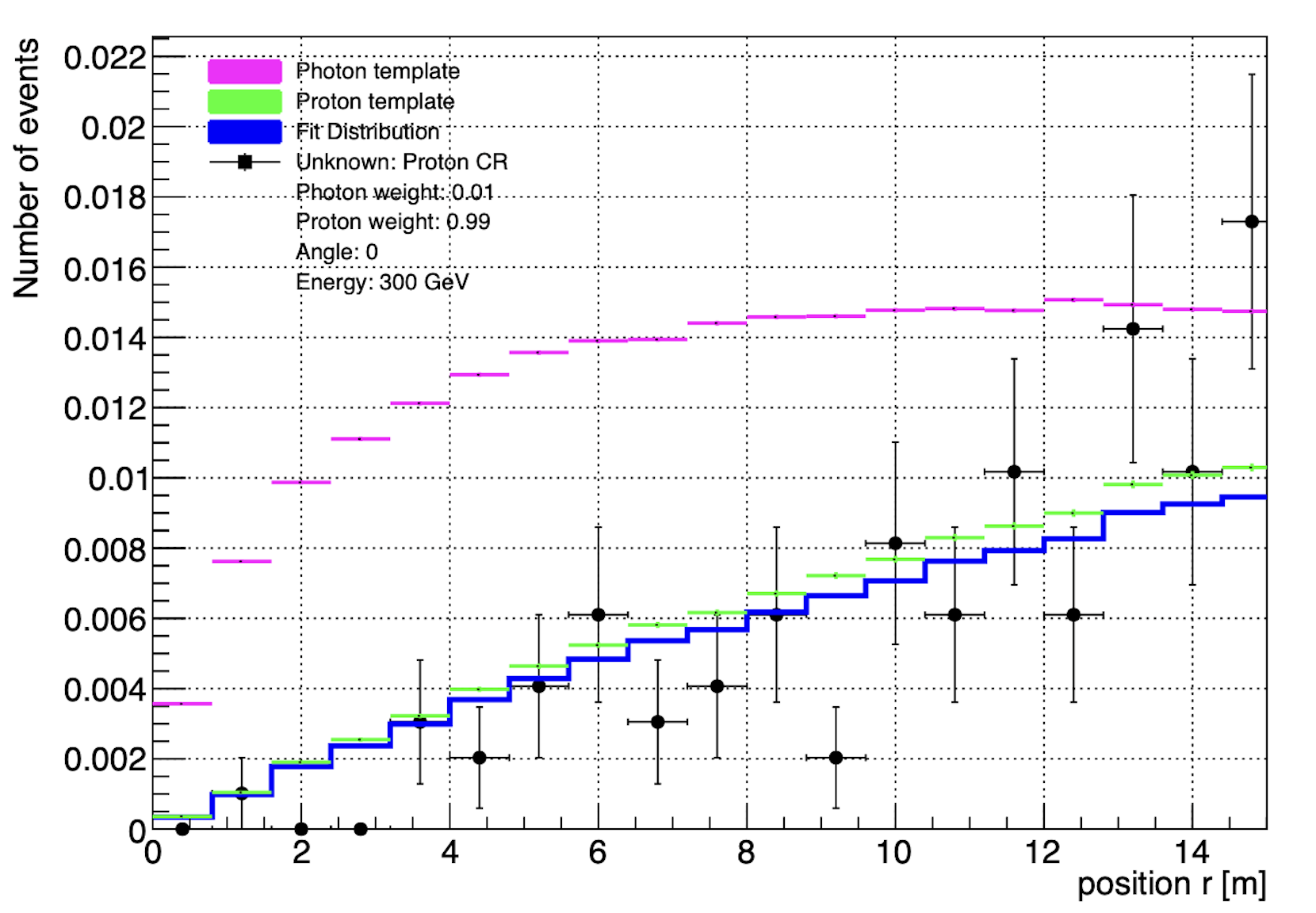}
        \caption{}
        \label{fig:subfig14}
    \end{subfigure}
    \caption{\small{
    Example of tagging for an unknown shower at an incidence angle of \(0^\circ\) and an energy of 300 GeV. (a) shows the fit to the Photon Monte Carlo template, while in (b), the plot shows the fit to the Proton Monte Carlo template. Black points represent the observed data for the unknown shower, and the blue line corresponds to the statistical fit. In this case, the Proton template yielded a statistical weight of 0.99, correctly classifying the shower as Proton-induced.
    }}
    \label{fig:single_angle_tagging}
\end{figure}

The combined distribution \(D_C\) is fitted to the observed distribution of an unknown shower to classify it as proton- or photon-induced as is shown in Figure~\ref{fig:single_angle_tagging}. During this process, the statistical weights \(W_p\) and \(W_\gamma\) are iteratively adjusted to minimize the residuals between \(D_C\) and the observed distribution. The blue line in each plot represents the statistical fit, while the black points correspond to the data of the unknown shower being evaluated. The closer the black points are to the blue line, the better the fit. The statistical weights \(W_p\) and \(W_\gamma\) are derived directly from this comparison.

Figure~\ref{fig:single_angle_tagging} illustrates the fits for the proton and photon templates applied to an unknown shower at \(0^\circ\) and 300 GeV.

\subsubsection{Efficiency, Purity, and Error Characterization}

It is essential to evaluate the accuracy of the tagging algorithm through quantitative measurements. To achieve this, we define the conditions under which tagging is successful. 
Focusing on proton-induced showers, the tagging algorithm can produce three possible outcomes:

\begin{itemize}
    \item Successful Tagging: \(W_p > 0.5\)
    \item Failed Tagging: \(W_p = 0.5\)
    \item False Positive Tagging: \(W_p < 0.5\)
\end{itemize}

A test dataset of 1,000 showers per angle across different energies was analyzed using these criteria. The tagging error percentage for each angle was calculated, considering both ``Failed Tagging'' and ``False Positive Tagging'' as error cases. 

False positives (FP) represent samples incorrectly classified as belonging to the photon category. To evaluate this, the False Positive Rate (FPR) is defined as:
\begin{equation}
    \text{FPR} = \frac{\text{FP}}{\text{N}_{\text{total}}}, \quad \text{where} \quad \text{N}_{\text{total}} = \text{FP} + \text{TP}.
\end{equation}

Here, \(\text{N}_{\text{total}}\) represents the total number of analyzed samples, including false positives (\(\text{FP}\)) and true positives (\(\text{TP}\)). The standard deviation of the FPR is calculated assuming a binomial distribution. To evaluate the efficiency (eff) and purity (pur) of the tagging algorithm, these metrics are defined as follows:

\begin{equation}
\text{eff} = \frac{\text{true selected}}{\text{true total}} \quad ; \quad \text{pur} = \frac{\text{true selected}}{\text{true selected} + \text{false selected}}
\end{equation}

Here, \(\text{true selected}\) represents the number of correctly tagged events, \(\text{true total}\) is the total number of true events, \(\text{false selected}\) is the number of misclassified events, and \(\text{true selected} + \text{false selected}\) represents the total number of events tagged. These metrics provide a quantitative framework for evaluating the tagger's performance in analyzing CR showers.

\begin{figure}[ht!]
    \centering
    \includegraphics[scale=0.5]{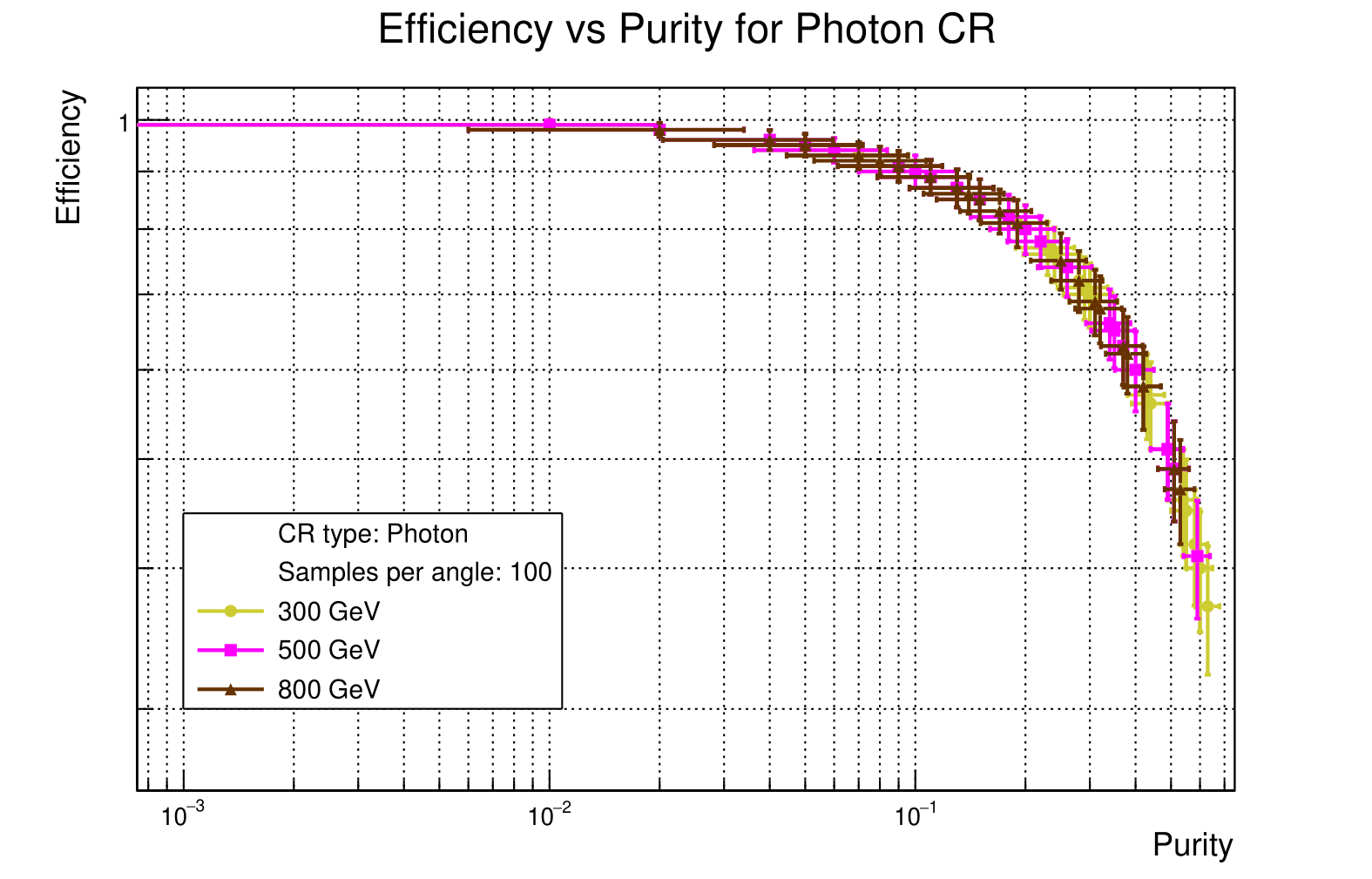}
    \caption{\small{
    Correlation between efficiency and purity for CR showers. This plot highlights the trade-off between the tagger's ability to correctly classify true events (efficiency) and its capacity to minimize misclassification of false events (purity).
    }}
    \label{fig:Efficiency_Purity_Ratio}
\end{figure}

When evaluating these metrics across all available angles generated through Monte Carlo simulations, the relationship between efficiency and purity are shown in Figure~\ref{fig:Efficiency_Purity_Ratio}. 
The results demonstrate a trade-off between the classifier's ability to correctly identify true events (efficiency) and its capacity to minimize misclassification of false events (purity). The analysis reveals that photon-induced showers exhibit high identification consistency. 

Although the initial description focuses on proton-induced showers, the tagging algorithm was applied to both gamma and proton simulations. Figures~\ref{fig:single_angle_tagging} and~\ref{fig:Efficiency_Purity_Ratio} demonstrate the separation power between the two classes, showing that even with a simple template-based approach, stable discrimination is achievable across energy and zenith angle.

The CONDOR tagging system demonstrates reliable performance in classifying CR showers, with photon-induced showers exhibiting higher identification consistency than proton-induced showers. However, limitations persist for showers at large incidence angles and low energies, where reduced particle density adversely impacts statistical fitting accuracy. Future improvements will focus on optimizing the tagging algorithm to address these challenges and enhance its robustness and applicability to real observational datasets by adding sources of backgrounds and noise. 

\subsection{CONDOR \texorpdfstring{$\gamma$}{gamma}-ray sensitivity}

\begin{figure}[ht!]
    \centering
    \includegraphics[scale=0.5]{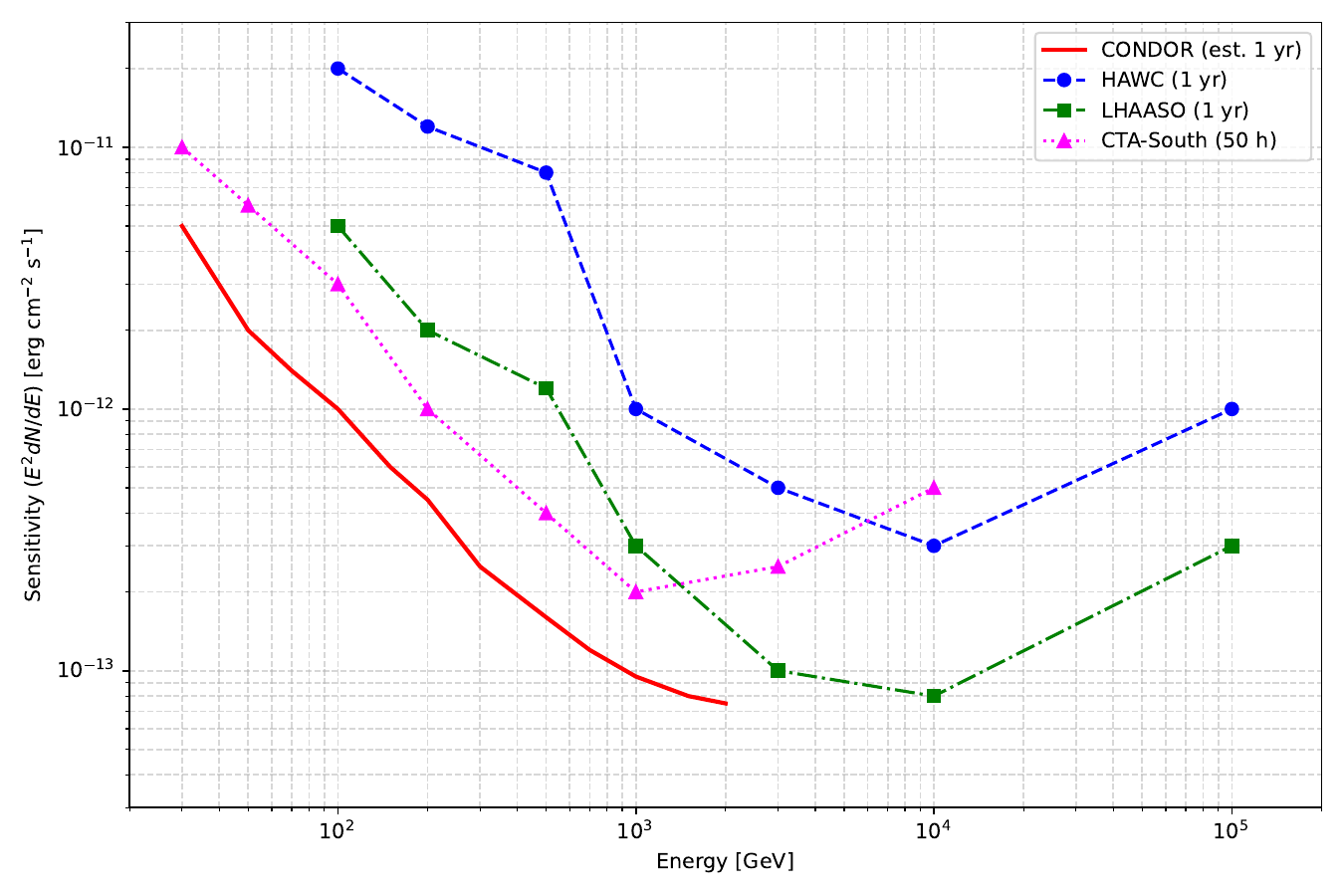}
    \caption{\small{Differential point-source sensitivity of the CONDOR $\gamma$-ray observatory for a one-year observation (90\% duty cycle), compared with the sensitivities of HAWC, LHAASO, and CTA-South for equivalent exposure times. CONDOR shows enhanced sensitivity in the sub-TeV range (~100–1000 GeV), complementing other observatories in this transitional energy domain.}}
    \label{fig:CONDOR_sense}
\end{figure}

Expected sensitivity of the CONDOR wide-field gamma-ray array for a one-year observation can be seen in Figure~\ref{fig:CONDOR_sense}. 
The sensitivity was estimated by simulating the detector response to $\gamma$-ray flux, using a standard background-limited approach. An effective area of 5700 m² is assumed, with angular resolution and gamma/hadron discrimination performance taken from full detector simulations. The dominant background is assumed to be cosmic-ray protons, and their residual contamination is estimated using the ROC curve from Figure~\ref{fig:Efficiency_Purity_Ratio}.
The sensitivity calculation includes a realistic trigger condition, requiring at least 0.2 GeV of energy deposition in a single detection module, along with a coincidence of at least two adjacent stations within a 50 ns time window. This sequential trigger logic achieves a high detection efficiency (>95\%) for gamma-induced showers, while significantly suppressing random background events.
A duty cycle of 90\% is assumed, consistent with continuous operation of ground-based air-shower arrays.
In the same figure, we show for comparison the projected sensitivities of HAWC, LHAASO, and CTA-South for similar exposure times (1 year for HAWC and LHAASO; 50 hours for CTA-South as a standard reference). As expected for its design, CONDOR shows superior sensitivity in the sub-TeV energy range (~100–1000 GeV), where its compact detector geometry and low-energy triggering allow efficient detection of low-energy air showers. In this regime, CONDOR complements existing instruments by filling the sensitivity gap between space-based telescopes (e.g., Fermi-LAT) and high-threshold ground-based observatories.
At higher energies (>10 TeV), instruments like LHAASO and CTA outperform CONDOR due to their larger instrumented areas and optimized reconstruction for high-energy showers. Nevertheless, CONDOR maintains competitive sensitivity in the TeV range, demonstrating its potential for wide-field, continuous monitoring of transient and steady gamma-ray sources.

\section{Summary and Discussion}
The CONDOR Observatory design represents a novel approach to high-altitude gamma-ray and CR detection, leveraging a dense scintillator array with a modest footprint optimized for a 100 GeV threshold. The design prioritizes modularity and robustness, ensuring efficient construction and deployment, as well as reliable operation under extreme environmental conditions. At 5,200 m above sea level, with a fill factor exceeding 90\% and precise timing synchronization provided by White Rabbit technology, CONDOR is uniquely positioned to detect low-energy cosmic and gamma rays. 

A key focus of this study has been the development and validation of angular reconstruction techniques, which leverage simulated air showers to determine the zenith angles of incoming CR. Using planar fitting methods, we demonstrated the feasibility of accurate angle reconstruction up to zenith angles of approximately 50$^{\circ}$ for both photon and proton showers. The precision of the reconstruction techniques is critical for distinguishing between CR types and understanding their spatial distribution.

Additionally, we implemented a Monte Carlo-based tagging system for gamma-ray and proton event discrimination. Our approach successfully classifies CR events with high efficiency and purity by employing statistical templates derived from simulated showers. The simulation results confirm that the observatory's high-altitude location significantly enhances particle detection rates by reducing atmospheric attenuation. This advantage translates into a more precise measurement of CR properties, complementing both satellite-based and ground-based detection efforts. The current CONDOR design does not include muon tagging capabilities, which are known to enhance gamma/hadron separation as demonstrated by LHAASO and SWGO. Incorporating a muon-sensitive layer or underground veto system remains an important option for future detector upgrades.

Future work will focus on refining angular reconstruction methodologies, optimizing gamma-proton discrimination techniques, and preparing for the construction and deployment phase of the CONDOR Observatory. 

We envision building the CONDOR Observatory as the world’s highest‑altitude gamma‑ray and cosmic‑ray observatory, distinguished by its exceptionally low energy threshold. Once completed, it will advance astroparticle physics and multimessenger astronomy~\cite{Astro2020} by measuring gamma rays in the 100 GeV--1 TeV range while operating continuously 24/7 in an all‑sky mode. In addition, it will complement other EAS gamma-ray observatories that target higher energies at lower altitudes~\cite{HAWK,ALPACA,SWGO,LHAASO} and Cherenkov imaging telescopes~\cite{CTAConsortium:2017dvg} that operate with narrower fields of view only at night.
\appendix

\acknowledgments
This work was funded by ANID PIA/APOYO AFB230003, Proyectos Internos de Investigación Multidisciplinarios USM 2024 PI$\_$M$\_$24$\_$02 and Millennium Institute of Subatomic Physics at High Energy Frontier: ICN2019$\_$044. The UC Riverside team work was supported by UC Riverside funding. We would like to thank the CLASS telescope crew for hosting us during our expedition and sharing their power and internet, the PAA personnel for assisting us during our visit, Miguel Mostaf\'a for sharing his wisdom and enthusiasm, and Matt Kauer and Chris Wendt for their various forms of support for our design inspired in IceTop. Also the authors thank the High Performance Computing Cluster (HPCC) at DFIS, UTFSM, San Joaquín, Santiago, Chile

We do not take the use of the CONDOR name lightly. We acknowledge that the Andean c\'ondor has been revered as sacred since time immemorial and serves as a national symbol for many nations in the Andean region, including Chile. We use the name respectfully, as a tribute to the peoples of the Andean region, in this project led by Chileans.


\bibliographystyle{JHEP}
\bibliography{bibliography}

\end{document}